\documentclass[acmtog,nonacm]{acmart}  

\AtBeginDocument{%
  \providecommand\BibTeX{{%
    \normalfont B\kern-0.5em{\scshape i\kern-0.25em b}\kern-0.8em\TeX}}}

\usepackage{dsfont}
\usepackage{wrapfig}
\usepackage{xspace}
\usepackage{amsmath}
\usepackage{enumitem}
\usepackage{algorithm}
\usepackage[noend]{algpseudocode}
\usepackage{xcolor}
\usepackage{multirow}
\usepackage{empheq}
\usepackage{mathtools}
\usepackage{float}
\usepackage{mdframed}
\usepackage{cleveref}
\usepackage[normalem]{ulem} 
\newmdenv[linecolor=gray, backgroundcolor=gray!10, roundcorner=5pt, skipabove=10pt, skipbelow=10pt]{graybox}

\usepackage[normalem]{ulem}   
\usepackage{xcolor}
\usepackage{etoolbox}         
\usepackage{cancel}           

\begin{document}

\title{\textsc{MiGumi}: Making Tightly Coupled Integral Joints Millable}

\author{Aditya Ganeshan}
\email{adityaganeshan@gmail.com}
\orcid{0000-0001-8615-741X} 
\affiliation{%
  \institution{Brown University}
  \city{Providence}
  \country{United States of America}
}

\author{Kurt Fleischer}
\email{kurt@pixar.com}
\orcid{0009-0007-1768-4591}
\affiliation{%
  \institution{Pixar Animation Studios}
  \city{San Francisco}
  \country{United States of America}
}

\author{Wenzel Jakob}
\email{wenzel.jakob@epfl.ch}
\orcid{0000-0002-6090-1121}
\affiliation{%
  \institution{École Polytechnique Fédérale de Lausanne (EPFL)}
  \city{Lausanne}
  \country{Switzerland}
}

\author{Ariel Shamir}
\email{arik@runi.ac.il}
\orcid{0000-0001-7082-7845}
\affiliation{%
  \institution{Reichman University}
  \city{Herzliya}
  \country{Israel}
}

\author{Daniel Ritchie}
\email{daniel_ritchie@brown.edu}
\orcid{0000-0002-8253-0069}
\affiliation{%
  \institution{Brown University}
  \city{Providence}
  \country{United States of America}
}

\author{Takeo Igarashi}
\email{takeo.igarashi@gmail.com}
\orcid{0000-0002-5495-6441}
\affiliation{%
  \institution{University of Tokyo}
  \city{Tokyo}
  \country{Japan}
}

\author{Maria Larsson}
\email{ma.ka.larsson@gmail.com}
\orcid{0000-0002-4375-473X}
\affiliation{%
  \institution{University of Tokyo}
  \city{Tokyo}
  \country{Japan}
}

\newcommand{\SXG}{\textsc{MXG}}
\newcommand{\SXGZero}{$\textsc{MXG}_{0}$}
\newcommand{\SXGR}{$\textsc{MXG}_{r}$}
\newcommand{\SXGFull}{\textsc{Millable Extrusion Geometry}}

\newcommand{\SurfaceTermFull}{\textsc{Surface Gap}\xspace}
\newcommand{\PathTermFull}{\textsc{Milling Path Distance}\xspace}
\newcommand{\SurfaceTerm}{$\mathcal{M}_{\text{S}}$}
\newcommand{\PathTerm}{$\mathcal{M}_{\text{P}}$}

\newcommand{\DataSize}{30}

\newcommand{\JointNameFull}{\textsc{Millable Kigumi}}
\newcommand{\JointName}{\textsc{MiGumi}}

\newcommand{\EFFull}{\emph{Millable Extrusion Field}}
\newcommand{\EF}{$\mathcal{E}$}

\newcommand{\aditya}[1]{\textcolor{blue}{[ADITYA: #1]}}
\newcommand{\dr}[1]{\textcolor{red}{[DANIEL: #1]}}
\newcommand{\maria}[1]{\textcolor{magenta}{[MARIA: #1]}}
\newcommand{\as}[1]{\textcolor{orange}{[ARIK: #1]}}

\newenvironment{packed_itemize}
{\begin{itemize}
    \vspace{-\topsep}
    \setlength{\itemsep}{1pt}
    \setlength{\parskip}{0pt}
    \setlength{\parsep}{0pt}
}{\end{itemize}}

\newenvironment{packed_enumerate}
{\begin{enumerate}
    \vspace{-\topsep}
    \setlength{\itemsep}{1pt}
    \setlength{\parskip}{0pt}
    \setlength{\parsep}{0pt}
}{\end{enumerate}}

\newtoggle{showchanges}
\togglefalse{showchanges}
\DeclareRobustCommand{\Add}[1]{%
  \iftoggle{showchanges}{\textcolor{blue}{#1}}{#1}%
}
\DeclareRobustCommand{\Del}[1]{%
  \iftoggle{showchanges}{\textcolor{red}{#1}}{}%
}
\DeclareRobustCommand{\Repl}[2]{
  \iftoggle{showchanges}{\Del{#2}\,\Add{#1}}{#1}%
}

\DeclareRobustCommand{\AddM}[1]{%
  \iftoggle{showchanges}{\color{blue}{#1}}{#1}%
}
\DeclareRobustCommand{\DelM}[1]{%
  \iftoggle{showchanges}{\begingroup\color{red}\cancel{#1}\endgroup}{}%
}
\DeclareRobustCommand{\ReplM}[2]{
  \iftoggle{showchanges}{\DelM{#2}\,\AddM{#1}}{#1}%
}

\DeclareRobustCommand{\Shep}[1]{%
  \iftoggle{showchanges}{\textcolor{magenta}{[Shepherd: #1]}}{}%
}

\begin{abstract}

Traditional integral wood joints, despite their strength, durability, and elegance, remain rare in modern workflows due to the cost and difficulty of manual fabrication.
CNC milling offers a scalable alternative, but directly milling traditional joints often fails to produce functional results
because milling induces geometric deviations---such as rounded inner corners---that alter the target geometries of the parts. Since joints rely on tightly fitting surfaces, such deviations introduce gaps or overlaps that undermine fit or block assembly.
We propose to overcome this problem by
(1) designing a language that represent millable geometry, and (2) co-optimizing part geometries to restore coupling.
We introduce \SXGFull\ (\SXG), a language for representing geometry as the outcome of milling operations performed with flat-end drill bits. \SXG\ represents each operation as a subtractive extrusion volume defined by a tool direction and drill radius. 
This parameterization enables the modeling of artifact-free geometry under an idealized zero-radius drill bit, matching traditional joint designs. Increasing the radius then reveals milling-induced deviations, which compromise the integrity of the joint.
To restore coupling, we formalize tight coupling in terms of both surface proximity and proximity constraints on the mill-bit paths associated with mating surfaces. We then derive two tractable, differentiable losses that enable efficient optimization of joint geometry.
We evaluate our method on \DataSize~traditional joint designs, demonstrating that it produces CNC-compatible, tightly fitting joints that approximates the original geometry.
By reinterpreting traditional joints for CNC workflows, we continue the evolution of this heritage craft and help ensure its relevance in future making practices.

\end{abstract}

\begin{teaserfigure}
  \includegraphics[width=\textwidth]{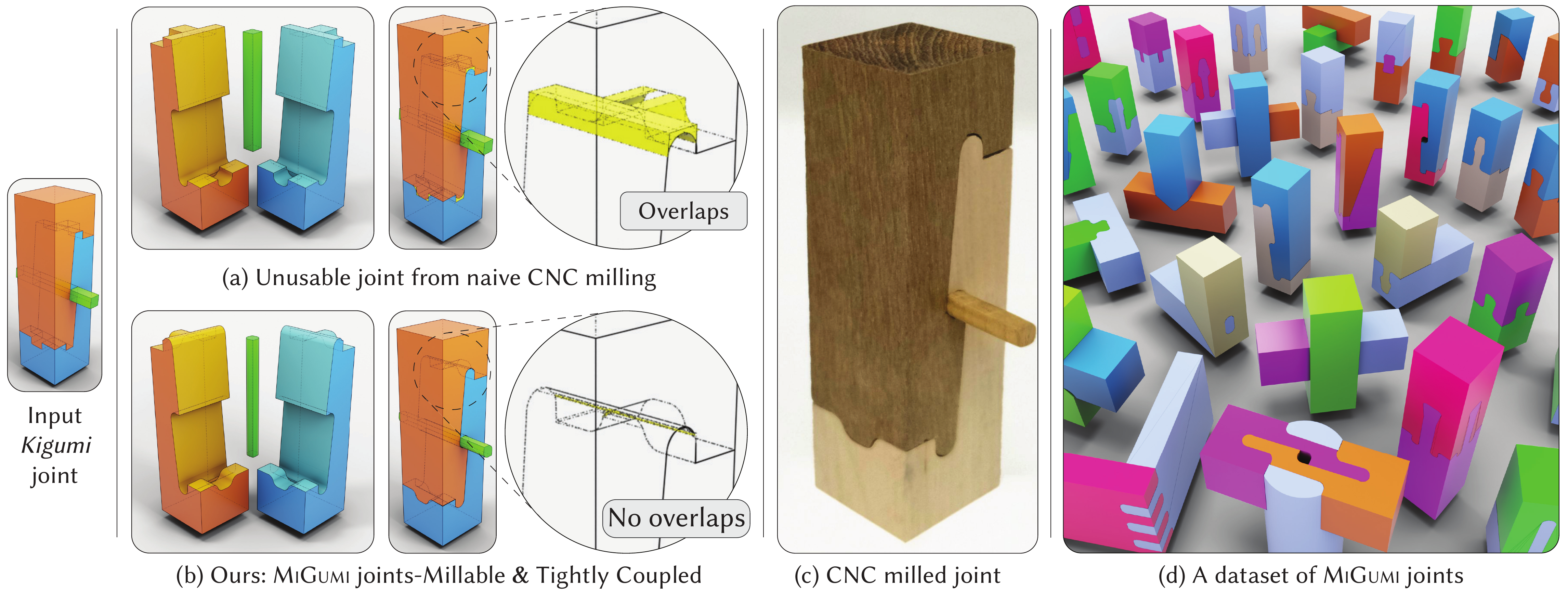}
  \caption{
  We present a method for fabricating integral joints using standard CNC machines. Nonzero-radius milling tools cannot produce sharp internal corners, and naively milling traditional designs leads to geometric deviations that make joints unassemblable. We introduce an optimization that adjusts geometry to yield millable, tightly coupled joints, and validate our approach on a dataset of \DataSize~ traditional designs.
  }
  \label{fig:teaser}
\end{teaserfigure}

\maketitle

\section{Introduction}

Integral joints (referred to as \textit{Kigumi} in Japanese) are assembled without glue, nails, or screws. Such joints have long been valued for their strength, reusability, and elegance. Developed across diverse woodworking traditions~\cite{zwerger2015wood}, they achieve mechanical function through tightly mating surfaces that constrain motion and transfer load. This property enables a rich design vocabulary, from sliding locks to hidden seams and stress-distributing features. Yet despite their utility, integral joints remain rare in modern manufacturing workflows. Producing them by hand requires high precision, specialized tools, and years of training—making them impractical in most contemporary workflows.

Three-axis CNC milling offers a compelling alternative. It is accessible, programmable, and widely used in modern woodworking. 
However, traditional integral joint designs contain sharp internal corners—features easily produced with hand tools but incompatible with cylindrical mill bits, which will inevitably cause rounding errors of aligned inner corners.
While such tool-induced artifacts are acceptable in many fabrication settings, integral joints are an exception. Their function depends entirely on tight surface contact between mating parts. Even small rounding errors can introduce gaps or overlaps that disrupt alignment, compromise fit, and in many cases, make assembly physically impossible.

\newcommand{\wjadded}[1]{#1}
\newcommand{\wjremoved}[1]{}

\begin{figure}[t]
    \centering
    \includegraphics[width=1.0\linewidth]{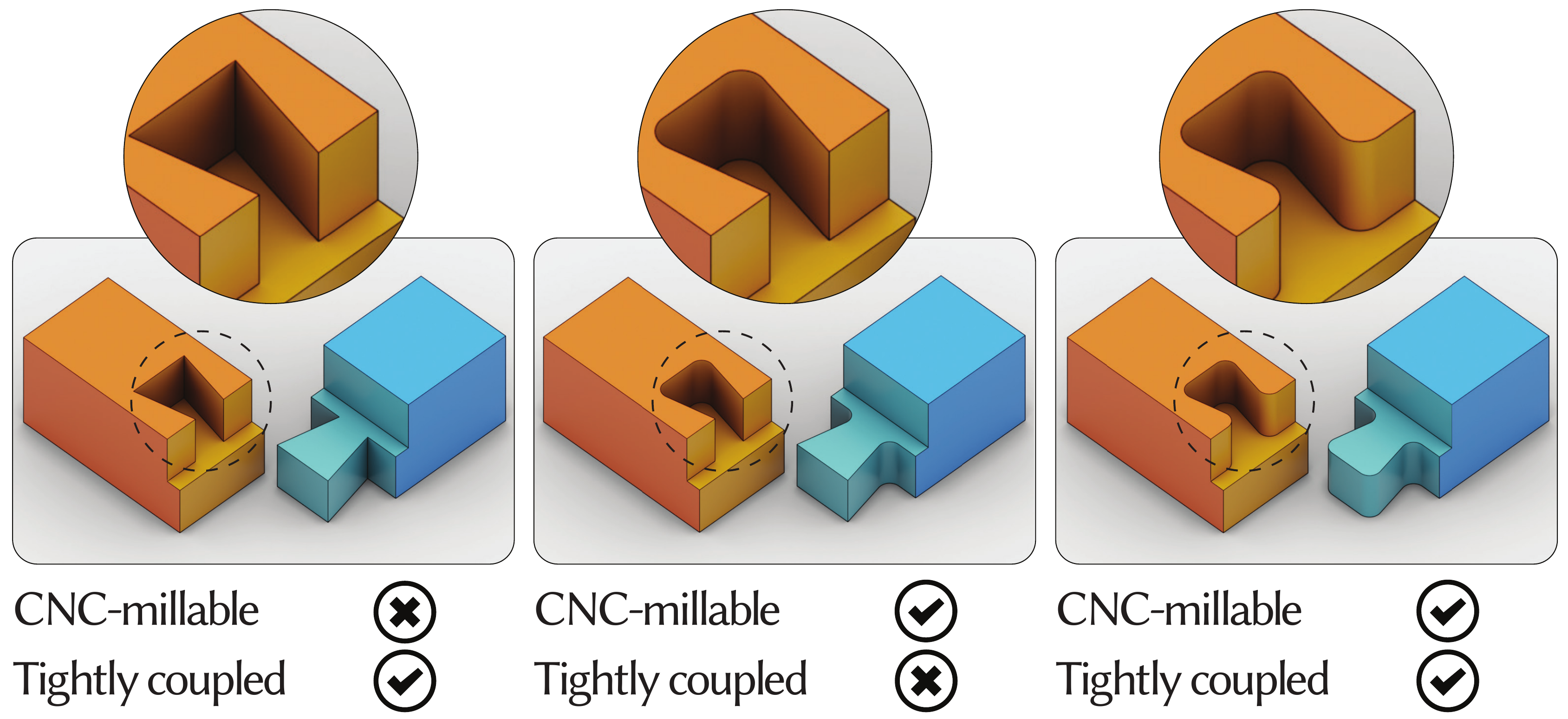}
    \caption{
    The path towards a \Add{millable \&} usable dovetail joint. (a) The original joint geometry contains features that cannot be milled as they are incompatible with the CNC machine's finite tool radius. (b) Individually adapting each part to be millable introduces overlaps that prevent successful assembly. (c) Our method jointly optimizes all contact surfaces to produce geometry that is both millable and tightly coupled.}    \label{fig:introduction:illustration-of-problem}
\end{figure}

Adapting integral-joint designs for CNC fabrication requires more than simply translating shapes into toolpaths: the geometry must be modified to anticipate milling artifacts and preserve precise contact between parts.
Figure~\ref{fig:introduction:illustration-of-problem} shows this issue on a simple dovetail joint: starting from a traditional design, naive milling introduces artifacts that break tight coupling, whereas our method produces geometry that is both millable and tightly coupled.
Most joints, however, are far more intricate: surfaces are often shaped by multiple milling passes from different directions, may interact with others cut on orthogonal faces, and many joints involve more than two parts (see Figure~\ref{fig:teaser}).  
Adapting such designs requires careful reasoning about how machining errors accumulate across interacting surfaces.  
We propose a tailored geometric representation and optimization procedure that accounts for these interactions and yields millable, tightly coupled joints across a broad range of joint designs.

Our first contribution is Millable Extrusion Geometry (MXG), a geometric language for parts that can be fabricated by a flat-end cylindrical milling bit. MXG models each part as a sequence of subtractions from an initial material stock (e.g., a wood block). Each subtraction operation removes a perpendicular 3D extrusion of a planar 2D shape expressed using a simple constructive solid geometry (CSG) language, and successive subtractions remove material along different extrusion directions. The representation is millable by design and parameterized by the tool radius, making it natural for analysis and mitigation of tool radius-related artifacts \Add{such as the unavoidable rounding of interior corners (cf. Figure~\ref{fig:introduction:illustration-of-problem})}. While MXG only covers a subset of geometry producible by CNC milling, we designed it to be expressive enough to handle large classes of integral wood joints. Its constrained structure enables key algorithmic simplifications that make our method \mbox{computationally efficient.}

Our second contribution is an optimization procedure based on two measures of tight coupling. The first, \SurfaceTermFull, penalizes separation between opposing surfaces to maintain contact after milling. The second, \PathTermFull, analyzes the milling path on both sides of a contact surface and constrains their closest-point distance to twice the tool radius, which complements the surface-based measure and stabilizes the optimization. Together, they provide a continuous and differentiable assessment of coupling under milling constraints, enabling robust gradient-based optimization of multi-part assemblies.

We begin with \SXG~programs authored in an idealized setting with zero tool radius and progressively increase this radius, adjusting the part geometry through optimization to maintain tight coupling. Because contact depends on the interaction of all parts, the optimization must be performed jointly. As we will show later, the structure of \SXG~and the design of our coupling metrics can be exploited to collapse the full 3D problem domain to a sparse set of planar 1D curves, which makes such an approach computationally feasible. The result is \JointNameFull~or \JointName~joints---\textit{millable} reinterpretations of traditional \textit{tightly coupled} \mbox{\JointName~integral joints. }

We evaluate our method on a dataset of \DataSize~traditional joint designs, modeled in \SXG~under an idealized zero-radius setting. Since no existing method directly addresses the problem of restoring tight coupling under milling constraints, we compare against two non-optimization-based variants. We show that these alternatives compromise either millability or coupling, while our approach consistently preserves both. Beyond quantitative analysis of fabrication validity, surface alignment, and geometric fidelity, we physically fabricate 9 joints using a standard 3-axis CNC machine—demonstrating that our results hold in real-world settings. Finally, we show that our method enables structured design exploration under milling constraints, significantly expanding the space of tightly coupled joint geometries that can be realized with modern CNC tools.

By integrating modern fabrication technologies with traditional joinery knowledge, our research contributes to reducing manual labor in production, expanding access to advanced joint design, and supporting the continued relevance of heritage woodworking techniques in contemporary manufacturing.

In summary, our contributions are:
\begin{packed_enumerate}
    \item \SXG, a DSL to model shapes as the outcome of milling operations with flat-end tools, enabling both  specification of artifact-free geometry under idealized conditions and systematic control over milling-induced artifacts.
    \item A differentiable optimization framework for restoring tight coupling under fabrication constraints by maximizing surface proximity and milling-path alignment between joint part \SXG\ programs.
    \item A dataset of \DataSize~traditional joint designs, modeled in \SXG, serving as a benchmark for millability-aware joint generation.
\end{packed_enumerate}

\noindent\textbf{Code and dataset available at:}~%
\href{https://bardofcodes.github.io/migumi}{\texttt{bardofcodes.github.io/migumi}}

\section{Related Work}

\subsection{Integral Joints \& Interlocking Assemblies} \label{subsection:related_works:joint_interlocking_assemblies}

\paragraph{Integral Joints.}
Several systems have proposed to support the design of \textit{external} joints that rely on separate connectors \cite{Magrisso2018, Kovacs2017} as well as \textit{integral} joints that are part of the components themselves \cite{decorativeJoinery2017, tsugite2020, finger_box_joints2014, matchsticks}. Within this latter area, \textit{Decorative Joinery}~\cite{decorativeJoinery2017} notably infers an internal joint geometry based on surface partitioning of an input object, enabling automatic generation of intricate joinery structures. 
However, the resulting designs are not constrained by CNC-milling limits and typically assume either 3D printing or manual carpentry for manufacturing the parts.

In contrast, systems have also been proposed for fabrication of integral joints with CNC milling~\cite{tsugite2020, overcut, suska_box_joint_2024}.
Some approaches~\cite{overcut} enable assemblability by introducing overcuts, but these come at the cost of reduced surface contact and often non-preferred appearance. 
\textit{Fingermaker!}~\cite{finger_box_joints2014} and tools within AutoDesk Fusion 360~\cite{suska_box_joint_2024} preserve tight contact but are limited to the relatively simple joint family of planar finger joints. 
Additional efforts have focused on curating fixed libraries of millable joint designs~\cite{gros, Kanasaki2013} without supporting new or parametrically-adjustable geometries, and often with overcuts. 
In contrast, our framework supports a broader class of joint types and enables continuous geometric variation while explicitly preserving tight coupling under realistic CNC constraints.

Most closely related to our work is \textit{Tsugite}~\cite{tsugite2020}, which focuses on joints that can be fabricated using 3-axis CNC milling, with emphasis on an interactive interface for rapid iteration and design feedback. 
Their modeling domain, however, is limited to low-resolution voxel grids (up to \(5 \times 5 \times 5\)), which restricts the range of representable geometries---making common designs such as dovetails fall outside their scope. 
Our work is inspired by this approach and aims to make integral joints \textit{millable} more broadly.

\paragraph{Interlocking Assemblies.}
A wide body of prior work has explored the computational design of interlocking assemblies, particularly in the context of rigid part furniture~\cite{interlockingFurniture2015, desia2018} and mechanical puzzles~\cite{highLevelInterlocking2022, interlockingAssemblies2019}.
We refer readers to a recent survey~\cite{sota_interlocking} for a comprehensive overview. 
While related, \emph{interlocking} is distinct from \emph{tight coupling}.
Interlocking assemblies allow local gaps as long as global motion is blocked; by contrast, integral joints often rely on precise surface mating for structural integrity and aesthetics.
Moreover, these methods typically do not integrate fabrication constraints, whereas our approach explicitly accounts for the geometric artifacts introduced by milling-based fabrication.

\paragraph{Other Properties of Integral Joints and Assemblies.}
Yet other work focus on other specific types of integral joints, such as those that are reconfigurable, meaning that parts connect in multiple different ways \cite{reconfigurableInterlocking2017, reconfigurableJoints2024} and cone joints, which are designed for simplifying the physical assembly process \cite{Wang2021MOCCA}.

\subsection{Geometry under Milling Constraints} \label{subsection:related_works:milling_constraints}

\paragraph{Retrofitting milling operations.}
Several systems model fabrication by retrofitting subtractive operations to efficiently match a target geometry~\cite{vdac2020, dscarver, Muntoni2018, curvature_milling, Zhang2025}. 
VDAC~\cite{vdac2020}, for instance, infers a sequence of milling actions that approximate a given shape, but does not account for fabrication-induced artifacts such as inner rounding or staircasing. 
This approach is suitable for small-scale objects fabricated on 5-axis CNC machines, where such artifacts can be resolved through secondary post-processing. Other research focuses on optimizing the CNC end bit geometry or toolpaths to minimize fabrication errors and to create smooth finishes~\cite{curvature_milling, Zhang2025}.
However, for deep structures---common in joinery---tool head reachability is limited, making artifact removal impractical. 
Instead, we optimize the design itself to preserve its functional property---tight coupling---despite unavoidable milling artifacts.

\citet{openingClosing2020} developed a general and efficient method for morphological \textit{opening} and \textit{closing} of geometries.
They demonstrate that the \textit{closing} operation can be used to generate tool-paths to cut 2D planar shapes. 
In contrast, our work focuses on 3D shapes fabricated with multiple milling operations where closing alone does not yield a feasible tool path.

Other work focus on tool path efficiency, reducing the milling time \cite{Wang2023}, which is beyond the target of our current study. Yet other work propose hybrid approaches, combining subtractive and additive manufacturing \cite{Zhong2023, Harabin2022}, which is also out of scope for this study.

\subsection{Other Related Directions} \label{subsection:related_works:other}
\paragraph{Optimizing functionality under fabrication constraints.}
There is a growing body of work that integrates fabrication constraints into optimization pipelines aimed at achieving structural or geometric goals within the domain-specific fabrication constraints of wire art \cite{Tojo2024Wireart}, grid shells, \cite{Becker2023GridShells}, board-based furniture \cite{umetani2012furniture, Noeckel2021}, laser cuts \cite{Abdullah2021}, and 3D-prints \cite{Martin2015crossSec}, to mention a few.
As for CNC-milling, while extensive efforts have been made to develop path planning algorithms for reproducing target geometries as closely as possible (refer to \cref{subsection:related_works:milling_constraints}), there is little research on adapting designs to fabrication constraints. 
An exception is \citet{MIRZENDEHDEL2020102825}, who incorporate accessibility fields into topology optimization for multi-axis machining.
In contrast to this work, however, our geometry is millable by construction, and the functional objective we optimize is not structural stiffness or load-bearing performance, but rather surface coupling fidelity between parts.

\paragraph{Alternative joint fabrication techniques.}
Beyond CNC milling, integral joints have been fabricated using a variety of processes, such as 3D printing~\cite{structCurves2024, decorativeJoinery2017, Luo2012, SONG2015137}, laser cutting~\cite{Baudisch2019, spring_fit, param_joint, fool_proof_joints}, and using power tools \cite{Leen2019}. Each fabrication technique has unique constraints and geometrical implications, so generality between techniques typically does not apply.

\begin{figure*}
    \centering
    \includegraphics[width=1.0\linewidth]{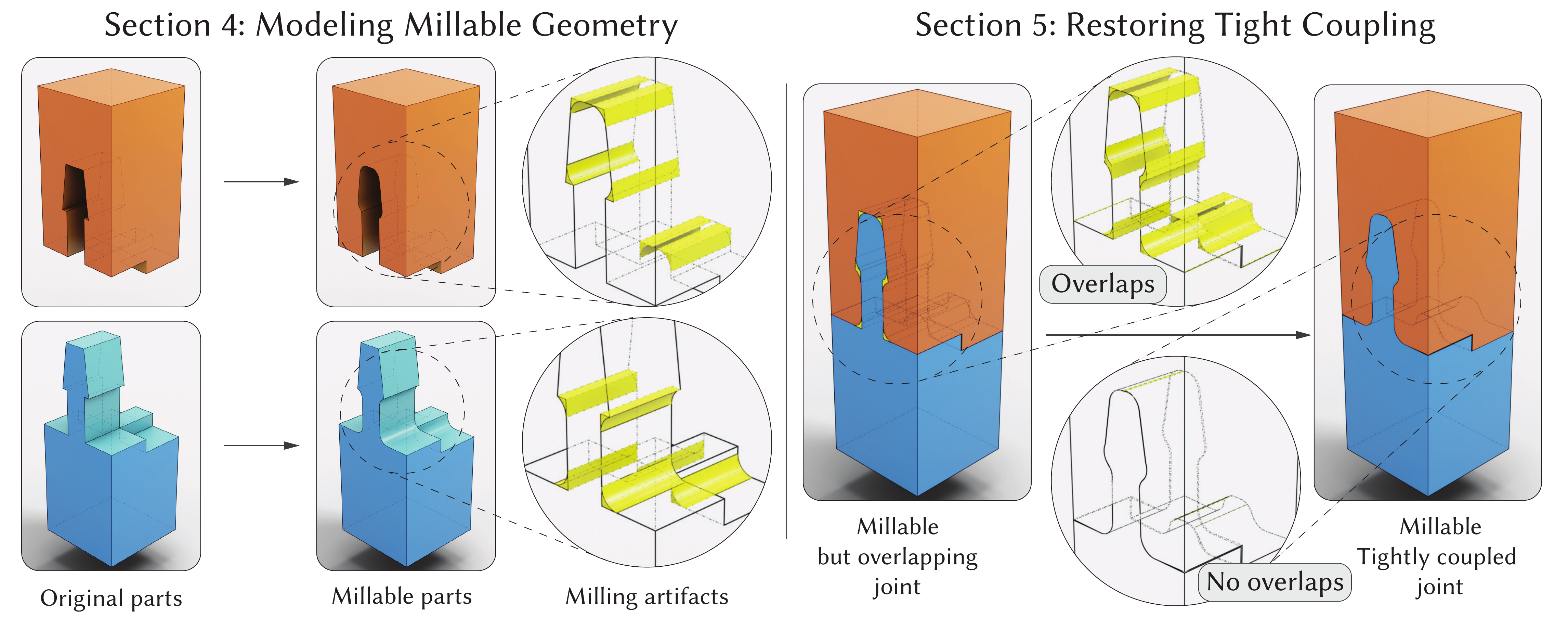}
    \caption{\textbf{Overview}: Our method addresses two challenges: modeling milling-induced distortions and restoring tight coupling.
Section~\ref{sec:millable_geom} introduces \SXG, a representation that models geometry as the outcome of milling operations, exposing rounding artifacts caused by milling.
Section~\ref{sec:tight_coupling} presents an optimization scheme that adapts joint geometry to restore coupling despite these artifacts.
    }
    \label{fig:overview}
\end{figure*}

\section{Modeling Millable Geometry}
\label{sec:millable_geom}

Milling-induced corner rounding artifacts break tight coupling between parts, introducing overlaps. Preserving coupling requires anticipating where artifacts occur and adapting the design accordingly.
We address this need with \SXGFull\ (\SXG), a representation that models geometry as the outcome of milling operations produced by flat-end, uniform-radius tools—a common setup in CNC woodworking.

\subsection{Preliminaries}
\label{subsec:prelims}

A millable solid \( P \subset \mathbb{R}^3 \) is represented as:
\begin{equation}
\label{eq:part_expr}
P = M - \bigcup_i V_i,
\end{equation}
where \( M \) is the material stock and each \( V_i \) is a volume removed by a milling operation. To ensure that \( V_i \) is physically realizable it must be expressible as a Minkowski sum\footnote{Given two sets $A, B \subset \mathbb{R}^n$, their Minkowski sum is defined as $A \oplus B = \{ a + b \mid a \in A,\, b \in B \}$.} with the tool’s swept volume. 
The rotational sweep of flat-end uniform-radius drill-bit yields a cylinder. Therefore, this can be written as:
\begin{equation}
\label{eq:mink-3d}
V_i = X_i \oplus \text{Cyl}_i,
\end{equation}
where $X_i$ is an arbitrary shape and $\text{Cyl}_i \subset \mathbb{R}^3$ is the rotational sweep of the drill bit for the $i$-th operation. 
To enforce accessibility, $\text{Cyl}_i$ is modeled as a semi-infinite cylinder extending away from the milling direction, ensuring that each $V_i$ is accessible from outside.

\begin{figure}
    \centering
    \includegraphics[width=1.0\linewidth]{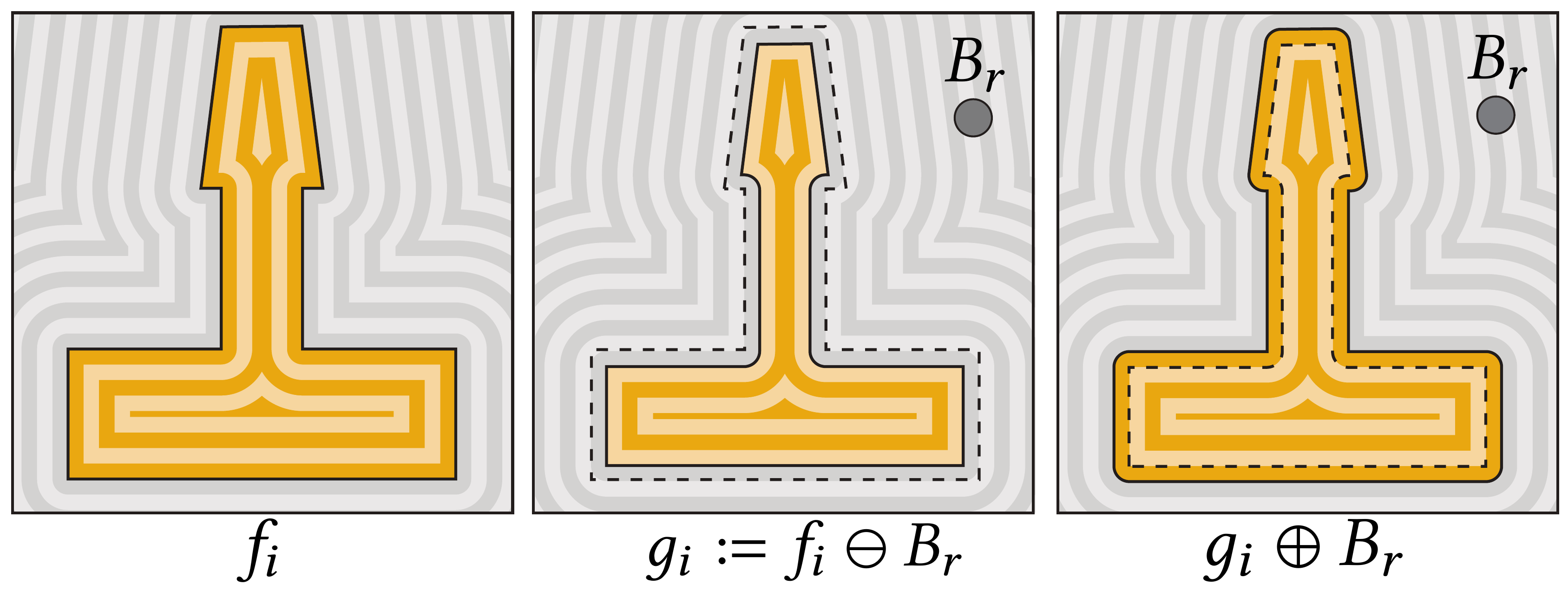}
    \caption{%
Morphological opening is the canonical way to derive geometry that can be safely removed by a milling tool. Given an implicitly defined input shape $f_i$ (left) and a structuring element $B_r$ representing the tool shape, the operation first applies a Minkowski subtraction (erosion, middle column) and then restores the removed volume by a subsequent addition (dilation, right column). The resulting shape excludes regions that the tool cannot reach and serves as a conservative, millable \mbox{approximation of the original geometry.}}
    \label{fig:morphological_opening}
\end{figure}

A constructive way to satisfy Eq.~\ref{eq:mink-3d} is to define \( V_i \) as the extrusion of a 2D region \( C_i \subset \mathbb{R}^2 \), embedded in a plane orthogonal to the milling direction \( \mathbf{n}_i \), over the semi-infinite interval \( (-\infty, h_i) \). 
If \( C_i \) satisfies the \textbf{Minkowski condition}, i.e., \( C_i = X_i \oplus B_r \) for some region \( X_i \) and disk \( B_r \) of radius \( r \), then the resulting volume \( V_i \) is millable by a flat-end tool of radius \( r \) (see supplementary for proof).  
\SXG\ encodes this construction through a primitive called the \EFFull. 
We now introduce this primitive and describes how it is composed to construct millable geometry by design.

\subsection{\SXGFull~(\SXG)}
\label{subsec:sxg}

We introduce \SXGFull~(\SXG), a representation for programmatically constructing millable geometry using explicitly parameterized subtractive primitives. 
The core construct in \SXG\ is the \EFFull~(\EF), which defines a single milling operation. Each \EF~generates a subtractive volume \( V_i \) that is guaranteed to be millable by a flat-end drill of radius \( r \).

\setlength{\intextsep}{0pt}       
\setlength{\columnsep}{6pt}       
\begin{wrapfigure}{r}{0.37\linewidth}
  \centering
  \includegraphics[width=1.0\linewidth]{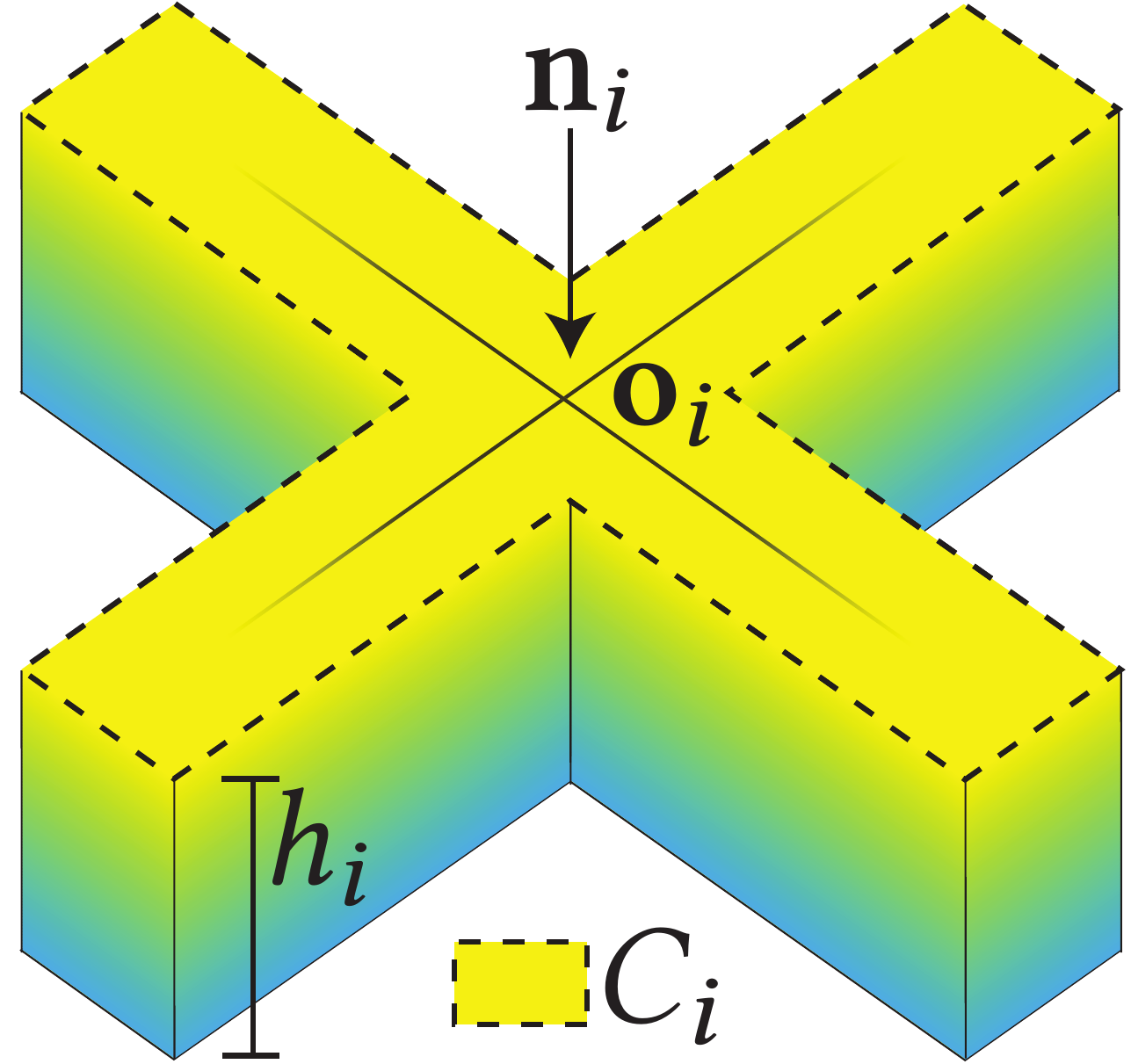}
  \caption{Extrusion with $r=0$}
  \label{fig:ef}
\end{wrapfigure}

An millable extrusion field \( \mathcal{E}_i \) is parameterized by a 2D signed distance function \( f_i : \mathbb{R}^2 \to \mathbb{R} \), an embedding plane \( p(\mathbf{o}_i, \mathbf{n}_i) \), an extrusion height \( h_i \), and a drill radius \( r_i \). 
It defines a volume \( \overline{\mathcal{E}_i} \) by extruding the sublevel set \( C_i = \{ x \mid f_i(x) \le 0 \} \) from \( -\infty \) to \( h_i \) along the direction \( \mathbf{n}_i \), as shown in Figure~\ref{fig:ef}.

To ensure millability, the base region $C_i$ must satisfy the Minkowski condition with respect to a disk of radius \( r_i \). We enforce this using morphological opening~\cite{SERRA1986283}: we erode \( f_i \) to obtain \( g_i = f_i \ominus B_{r_i} \), then dilate the result to define the updated region
\[
C_i^r = \{ x \mid (g_i \oplus B_{r_i})(x) \le 0 \}.
\]
This guarantees that the extruded volume is physically realizable with a drill of radius \( r_i \). The full extrusion is then given by \( \overline{\mathcal{E}_i^e} = \text{extrude}(C_i^r, \mathbf{n}_i, h_i) \).
Figure~\ref{fig:morphological_opening} illustrates this process.

\begin{figure}
    \centering
    \includegraphics[width=1.0\linewidth]{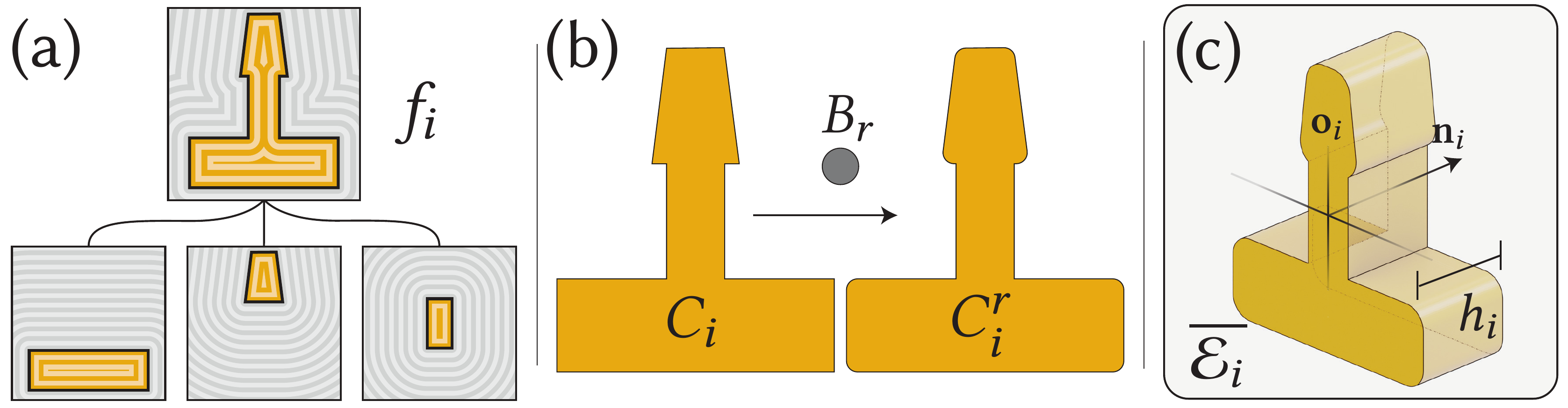}
    \caption{We construct a millable extrusion by (a) defining a 2D SDF \( f_i \) using a symbolic CSG tree, (b) applying morphological opening to obtain a millable base region \( C_i^r \), and (c) extruding this region along direction \( \mathbf{n}_i \) to produce the solid volume \( \overline{\mathcal{E}_i} \).
    }
    \label{fig:extrusion_simple}
\end{figure}

Figure~\ref{fig:extrusion_simple} shows the full construction pipeline: a 2D profile defined using symbolic CSG is morphologically opened and extruded to produce a valid subtractive volume. 
$f_i$ can be specified using an arbitrary constructive solid geometry (CSG) expression, allowing for rich and parameterized profiles.

\begin{figure}
    \centering
    \includegraphics[width=1.0\linewidth]{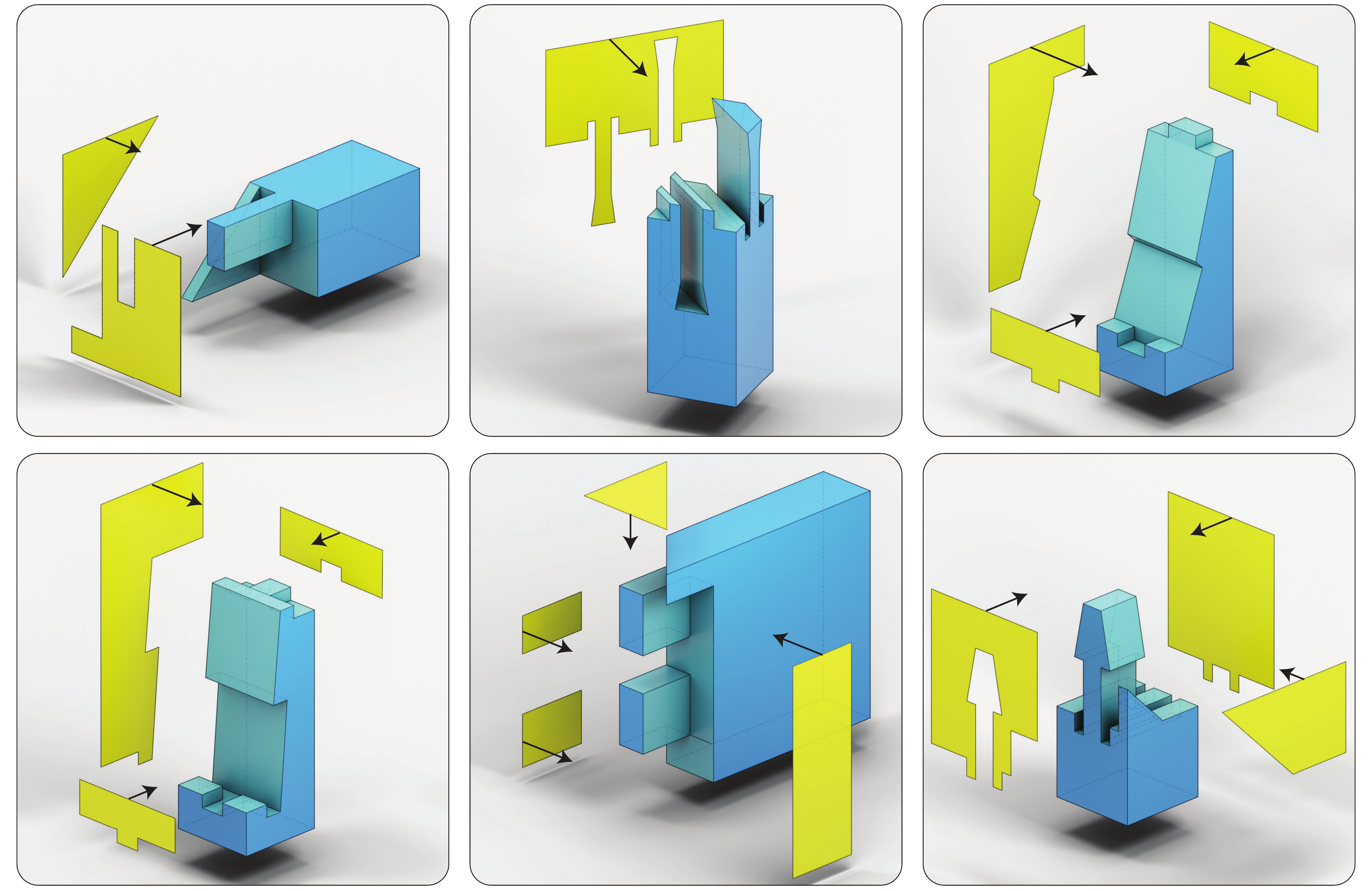}
    \caption{
Examples of traditional integral joint parts modeled using \SXG. 
Each example shows the resulting solid along with the milling directions and 2D contours of the subtractive extrusions used in its construction. 
These examples capture the original design but require a physically infeasible milling tool with radius $r=0$.
}
    \label{fig:sxg_program}
\end{figure}

\paragraph{Part Programs} \SXG~defines complete part geometries as a sequence of millable subtractions. Each part is represented as a material stock \( M \) minus a set of millable extrusion volumes, i.e., $P = M - \bigcup_i \overline{\mathcal{E}_i}$, as given in Equation~\ref{eq:part_expr}. This formulation ensures that the resulting geometry is physically millable by construction.
Figure~\ref{fig:sxg_program} illustrates the expressiveness of \SXG: we illustrate varied part geometries constructed by subtracting multiple different extrusions.

\paragraph{Modeling Milling-induced Artifacts}
\label{subsec:controlled_artifacts}

\begin{figure}
    \centering
    \includegraphics[width=1.0\linewidth]{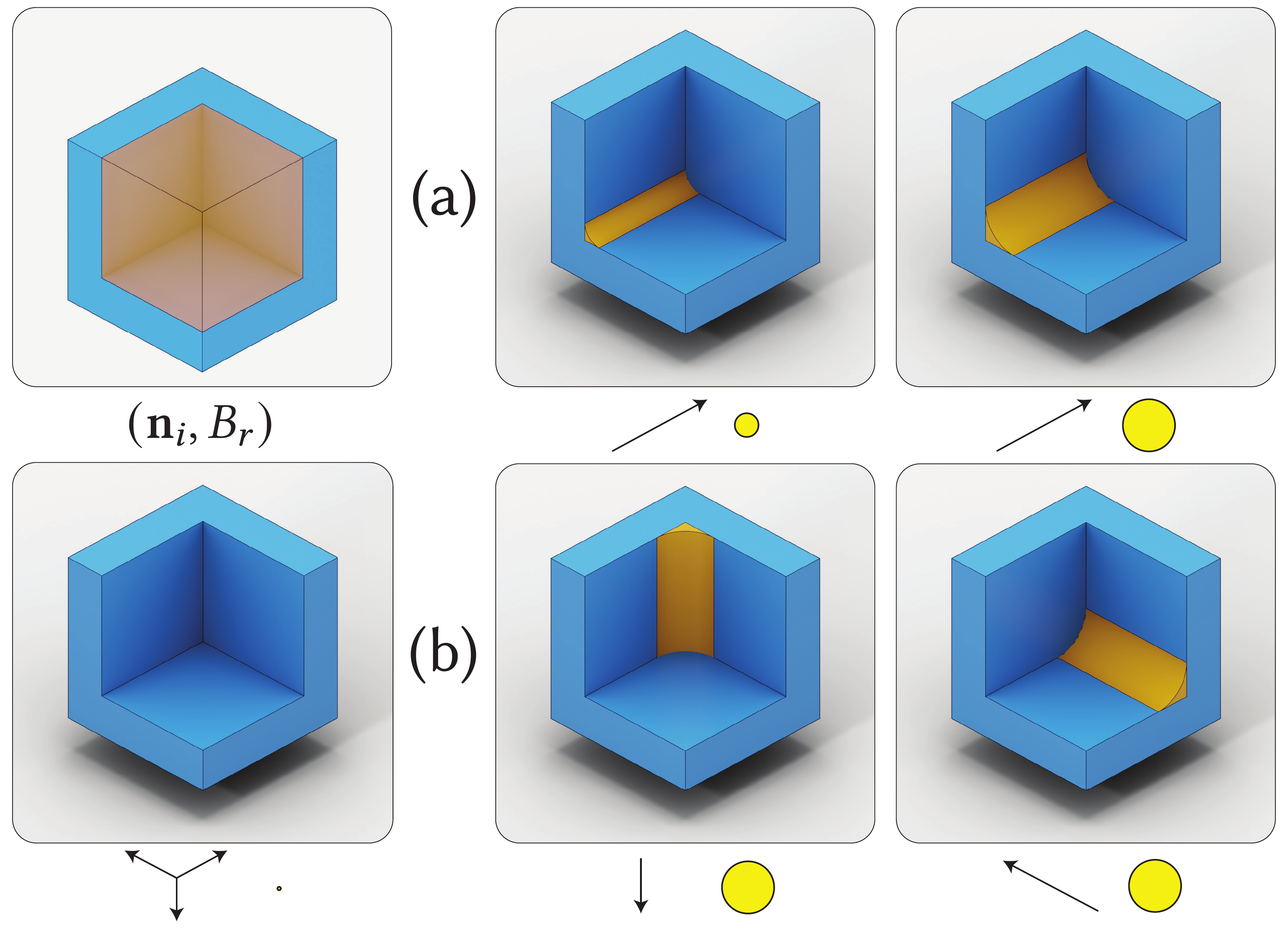}
    \caption{
Our \textsc{Millable Extrusion Geometry} (\SXG) representation enables explicit control over milling-induced artifacts via the drill radius $r$ and subtraction direction $\mathbf{n}_i$ parameters. The left column shows the target part with an ideal tool radius ($r=0$). (a) Increasing $r$ amplifies corner rounding. (b) Varying $\mathbf{n}_i$ shifts both the location and orientation of artifacts.}
    \label{fig:sxg_delta}
\end{figure}

A key feature of \SXG\ is that it provides explicit control over milling artifacts through per-subtraction parameters: the drill direction \( \mathbf{n}_i \) and radius \( r \). 
Figure~\ref{fig:sxg_delta} (a) illustrates how increasing the drill radius introduces progressively stronger corner rounding, while Figure~\ref{fig:sxg_delta} (b) shows how varying the milling direction alters the artifact—even when the underlying geometry remains identical at zero radius.

\paragraph{Modeling \SXGZero~programs}
To express clean, artifact-free geometry, we author joint designs under an idealized setting of a zero-radius mill bit $r=0$, yielding what we refer to as \SXGZero\ programs. Although not directly fabricable, these programs capture the intended shape of traditional joints. As we increase the tool radius $r$, evaluating the same program yields geometry distorted by milling-induced artifacts. \SXG\ thus provides a modeling framework that captures both idealized geometry and its distortion under fabrication constraints.

\begin{figure*}
    \centering
    \includegraphics[width=1.0\linewidth]{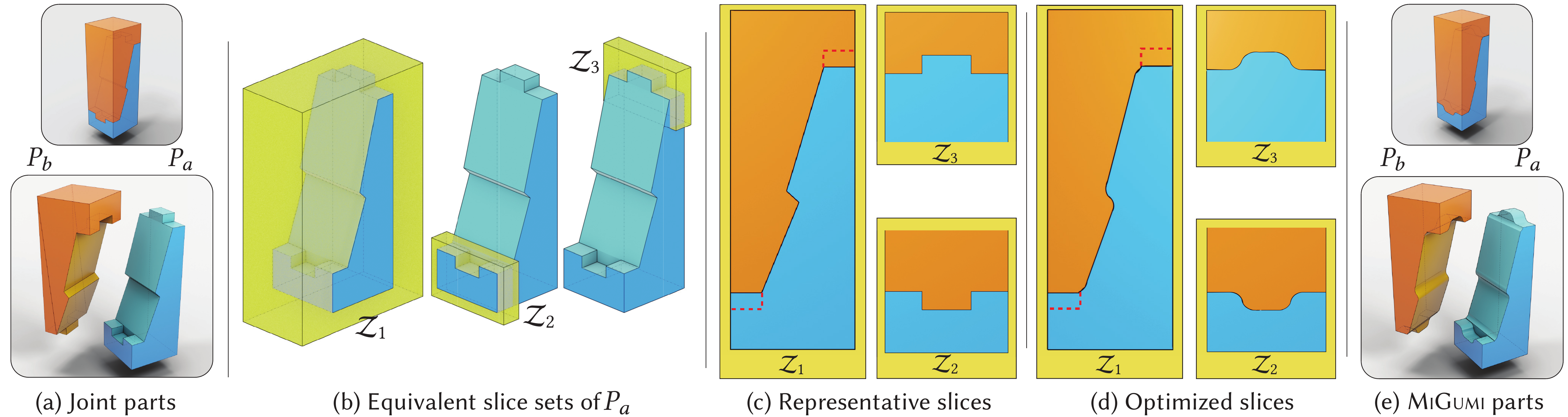}
    \caption{%
(a) Our optimization pipeline exploits a structural property of \SXG\ representation to greatly reduce the problem complexity. Based on its definition as a composition of subtractive planar extrusions, we observe that tight coupling can be fully characterized in 1D slices that lie perpendicular to the extrusion directions. (b) Many of the slices exhibit identical behavior and can be further grouped into a few representative planar sets. (c) The optimization domain then reduces to 1D planar curves within these representative slice sets. The interface of $\mathcal{Z}_1$'s slice with extrusions from other directions is shown as a  dashed red line. Such boundaries are kept fixed.
(d) We optimize this reduced space to compute tightly coupled geometry, and (e) reassemble the results into the final 3D MiGumi parts. 
    }
    \label{fig:planar_slices}
\end{figure*}

\section{Restoring Tight Coupling}
\label{sec:tight_coupling}

We now address the challenge of restoring tight coupling between parts in the presence of milling-induced artifacts. 
This requires reasoning about both the geometry of mating surfaces and the milling operations that generate them. 
We first formalize tight coupling using two complementary measures: one that evaluates surface contact between parts, and another that constrains the spacing between paired milling paths. 
These measures are differentiable and efficiently computable for \SXG~geometry, enabling their use as loss functions. 
We then describe an optimization process that takes \SXGZero~programs as input and updates their parameters to maintain coupling under a target milling radius $r$.

\subsection{\SurfaceTermFull}
\label{subsec:surface_term}

Let a joint system be composed of parts \( \mathbf{P} = \{P^1, \ldots, P^n\} \). 
As tight coupling is typically expected only in the interior region of the joint system, we define the \emph{coupling volume} \( \Omega \subset \mathbb{R}^3 \) as the region within which all surfaces must be in tight contact. 
In our approach, we simply define \( \Omega \) as the internal volume of the joint system excluding its exposed surfaces.

Let \( \mathbf{P} = \{P^1, \ldots, P^n\} \) be parts in \( \mathbb{R}^3 \), and let \( \Omega \subset \mathbb{R}^3 \) be the coupling volume.  
The \SurfaceTermFull~is defined as:
\begin{equation}
\label{eq:measure_sc}
\mathcal{M}_\text{S}(\mathbf{P}; \Omega) = 
\sum_{a=1}^n \int_{\partial P^a \cap \Omega} \min_{b \neq a} \mathcal{D}(x, P^b) \, dA(x),
\end{equation}
where $\partial P$ is the surface/boundary of part $P$,  $x$ is the center of an infinitesimal surface patch $dA(x)$ on the surface $\partial P^a$ and \( \mathcal{D}(x, P^b) \) is the Euclidean distance from point \( x \) to the surface of part \( P^b \).
We call a joint system \( \mathbf{P} \) \textit{tightly coupled} within \( \Omega \) if \( \mathcal{M}_\text{S}(\mathbf{P}; \Omega) = 0 \).
That is, every point on a surface within the coupling volume must be in exact contact with the surface of another part. 

\subsection{\SurfaceTermFull~ for \SXG~ Shapes}
\label{subsec:planar_obj}

When each part \( P \) is constructed using an \SXG\ program, a natural decomposition of its boundary surface emerges that allows us to reformulate the surface gap \( \mathcal{M}_\text{S}(P; \Omega) \) into a significantly more tractable form.

\begin{figure}
    \centering
    \includegraphics[width=1.0\linewidth]{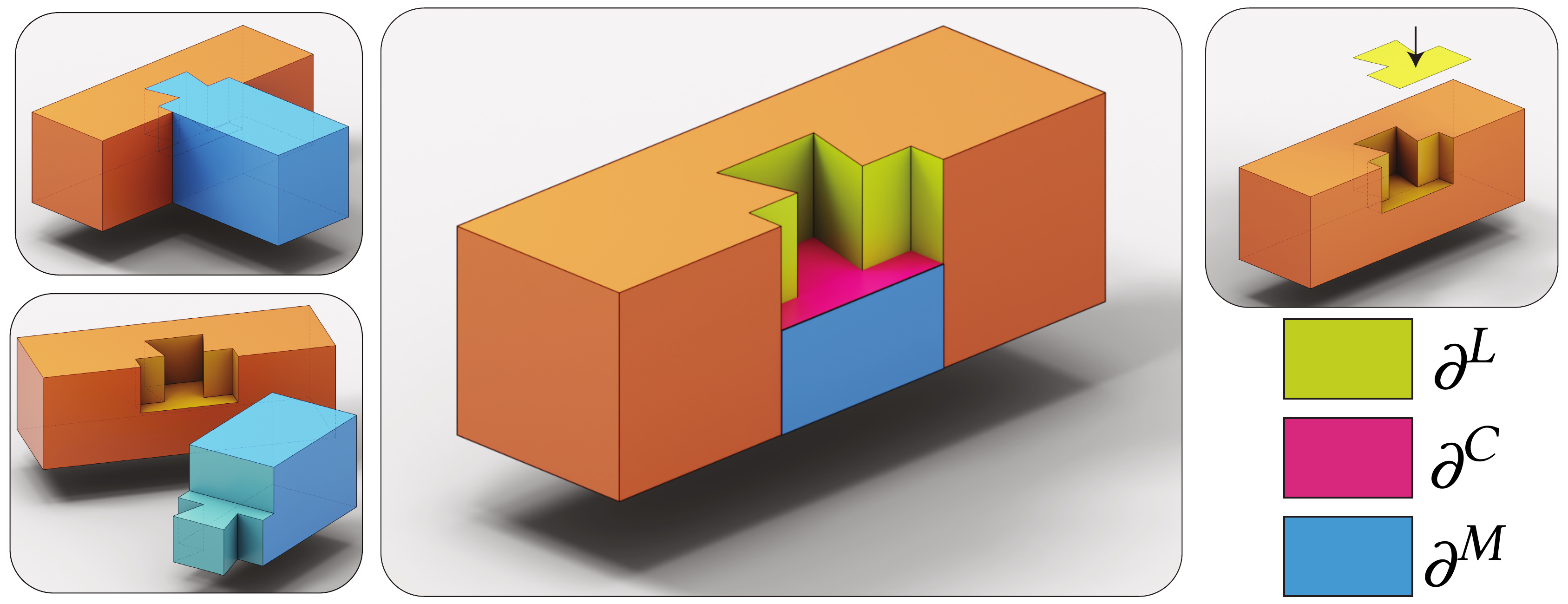}
    \caption{%
Our method incorporates a \SurfaceTermFull\ measure that promotes tightly coupled part interfaces. Its efficient evaluation relies on decomposing each part’s boundary into material surfaces ($\partial^M$), cap surfaces ($\partial^C$), and lateral surfaces ($\partial^L$), illustrated here using a dovetail joint.
    }
    \label{fig:surface_partitions}
\end{figure}

Recall from Section~\ref{subsec:sxg} that a part $P$ is expressed as \( P = M - \bigcup_i \overline{\mathcal{E}_i} \), where each \( \overline{\mathcal{E}_i} \) denotes the volume removed by extruding a planar region \( C_i \) along direction \( \mathbf{n}_i \), starting from plane \( p(\mathbf{o}_i, \mathbf{n}_i) \) and extending over the interval \( (-\infty, h_i) \).
As a result, the surface of part \( P \) is composed of three disjoint subsets (Figure~\ref{fig:surface_partitions}): the portion inherited from the material stock: \( \partial^M = \partial P \cap \partial M \), the \emph{lateral surfaces} formed by the sides of each extrusion, denoted by \( \partial_i^L \), and the \emph{cap surfaces} formed by the terminal flat ends of each extrusion, denoted by \( \partial_i^C \). 
Using this decomposition, \SurfaceTerm~for a single part $P$ can be rewritten as:
\begin{equation}
\begin{gathered}
\label{eq:gap-decomposition}
\mathcal{M}_\text{S}(P; \Omega) =
\sum_i \left( \int_{\partial_i^L} \mathcal{T} +  \int_{\partial_i^C} \mathcal{T} \right) + \int_{\partial^M} \mathcal{T}, \\
\text{where } \mathcal{T} = \min_{P^b \neq P} \mathcal{D}(x, P^b) \, dA(x).
\end{gathered}
\end{equation}

\begin{figure*}
    \centering
    \includegraphics[width=1.0\linewidth]{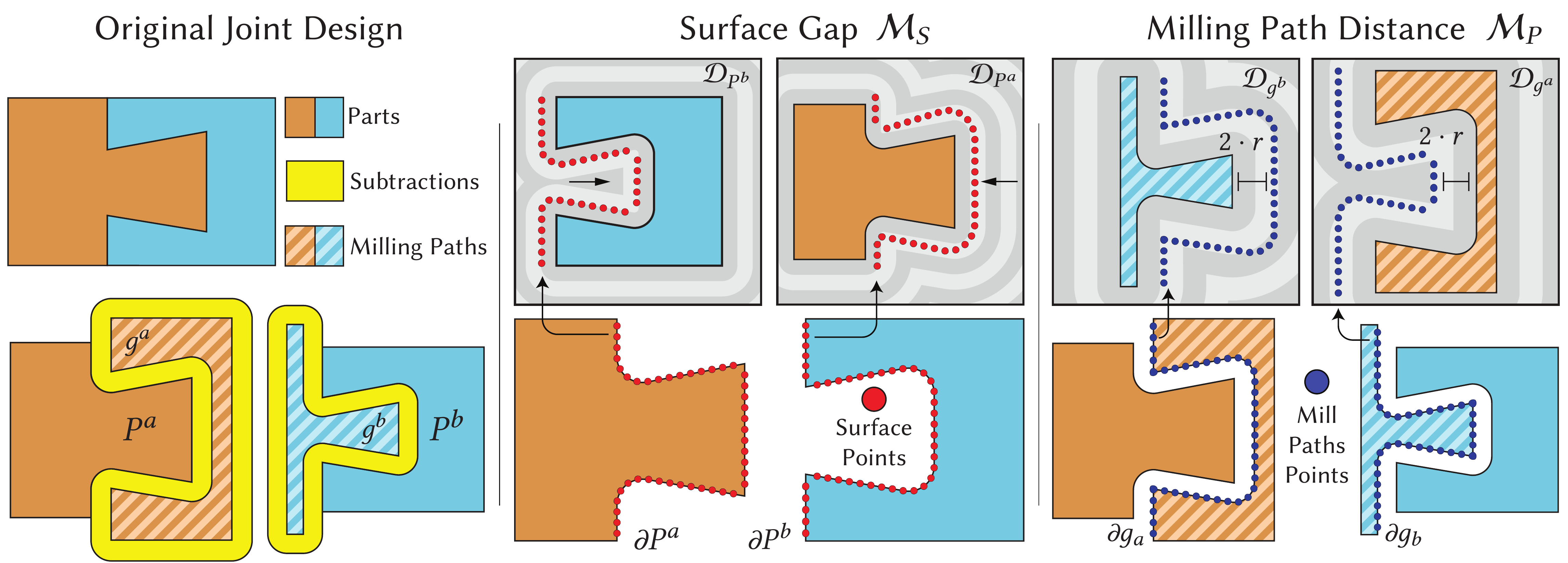}
    \caption{%
We optimize two complementary measures to ensure tight coupling: the \SurfaceTermFull\ evaluates the surface contact of adjacent part pairs, while the \PathTermFull\ constraints the spacing between their milling paths. Left: given an input joint design, we first convert it into tool paths (striped areas) with a nonzero tool radius. These initial paths do not yet yield tightly coupled geometry. Middle: the \SurfaceTermFull\ measures the proximity of opposing surfaces by numerically integrating the closest-point distances measured from a discrete sampling (red points) along the underlying plane curve within a representative slice (cf.~Figure~\ref{fig:planar_slices}). Right: the \PathTermFull\ evaluates the distance between milling paths by inserting one path into the planar signed distance field (SDF) of the opposing part and smoothly penalizing values that deviate from the desired distance ($2\cdot r$). We optimize both via gradient-based methods.}
    \label{fig:losses}
\end{figure*}
For clarity, we omit intersection with the coupling volume \( \Omega \), though all integrals are restricted to surface regions within it. Note that the surfaces $\partial^L_i$ and $\partial^C_i$ for an extrusion $\mathcal{E}_i$ is formed only on regions where material exists (i.e. the part volume remaining after subtracting other extrusions).

We now focus on the lateral surface term \( \partial_i^L \). Since this surface is generated by sweeping a 2D profile \( C_i \) along direction \( \mathbf{n}_i \), we can partition it into infinitesimally thin slices orthogonal to \( \mathbf{n}_i \), each corresponding to a planar curve. This allows us to express the 3D surface integral as a 1D integral over a family of planar contours:
\begin{equation}
\int_{\partial_i^L} \mathcal{T} =  
\int_{z=-\infty}^{h_i} \left( 
\int_{\partial C_i(z)} \min_{P^b \neq P} \mathcal{D}(x, P^b) \, ds(x) 
\right) dz,
\end{equation}
where \( \partial C_i(z) = \partial C_i \cap P(z) \) denotes the boundary of the 2D extrusion profile \( C_i \) that creates a surface on part \( P \) at depth \( z \) along the milling direction.

Now, rather than measuring the 3D distance from each surface point \( x \in \partial C_i(z) \) to the full surface of \( P^b \), we compute the 2D distance to its planar intersection at the same slice height \( z \), denoted \( P^b(z) \). 
Note that this yields a stronger criterion: the 2D clearance upper bounds the true 3D clearance, ensuring that any improvement under this measure also improves the true surface contact.
Since each slice has infinitesimal thickness, this effectively measures the gap between projected part contours in 2D.
A key observation now follows: for any two slicing planes \( z_1 \) and \( z_2 \), the corresponding slice integrals are identical whenever (i) the domain of integration \( \partial C_i(z) \) remains the same, and (ii) the set of projected contours \( \{ P^b(z) \}_{b \neq P} \) from other parts does not change. In such cases, the integral can be evaluated once on a representative slice and scaled by the size of the equivalent slice set.

This insight significantly reduces computational cost: surface gap for the lateral surfaces can be evaluated by identifying the set of  equivalent slices, computing a single 2D contour integral for each, and summing their contributions:
\begin{equation}
\label{eq:equivalence_class_gap}
\int_{\partial_i^L} \mathcal{T} 
\approx \sum_{k=1}^m w_k \cdot 
\left( 
\int_{\partial C_i(z_k)} 
\min_{P^b \neq P} \mathcal{D}(x, P^b(z_k)) \, ds(x) 
\right),
\end{equation}
where \( w_k \) is the size of the \( k \)-th equivalence class, and \( z_k \) is any representative height from that class.
Figure~\ref{fig:planar_slices}(a–c) illustrates our planar slice-based evaluation. 
(a) shows a two-part joint, (b) depicts the three planar equivalence slice sets for one part, and (c) highlights one representative slice from each set. 
Together, the contours \( \partial C_i(z_k) \) on each representative slice span the lateral surface within the coupling volume \( \Omega \), and are used to efficiently estimate surface gap.
This approach is especially effective for traditional integral joints, where large regions often satisfy the equivalence condition—allowing surface gap to be estimated efficiently from a small number of representative contours.

To keep the set of representative slices compact, slices are grouped along each extrusion directions after subtracting other extrusions whose directions are not parallel with the current one. Further, during optimization, we keep the contours along these inter-direction interfaces fixed, preventing shifts that would otherwise change their equivalence class. The red dashed lines in Figure~\ref{fig:planar_slices}(c-d) illustrates such a case.

Critically, evaluating surface gap on this compact set of planar slices not only improves efficiency but also suffices to preserve global coupling. If the initial configuration is tightly coupled and only the extrusion profiles \( f_i \) are modified, then enforcing zero deviation on these slices ensures that the overall surface gap \( \mathcal{M}_\text{S}(\mathbf{P}; \Omega) \) remains zero, i.e., contributions from cap surfaces  and material surfaces  also evaluates to zero.

\subsection{\PathTermFull}

Recall that each extruded region \( C_i \) is formed by dilating a base curve \( g_i \) by a disk of radius \( r_i \), i.e., \( C_i = g_i \oplus B_{r_i} \). 
We assume $g_i$ is an exact signed distance function, and therefore the Minkowski Sum can be performed using the dilation operator.
The boundary \( \partial C_i \) defines the visible surface of the milled region, while the zero-level set of \( g_i \) can be interpreted as the tool path that generates \( \partial C_i \). We refer to this zero contour \( \partial g^i := \{x \mid g_i(x) = 0\} \) as the mill path associated with extrusion \( \mathcal{E}_i \). 
Note that this is not the tool path used for fabrication.

\begin{figure}
    \centering
    \includegraphics[width=1.0\linewidth]{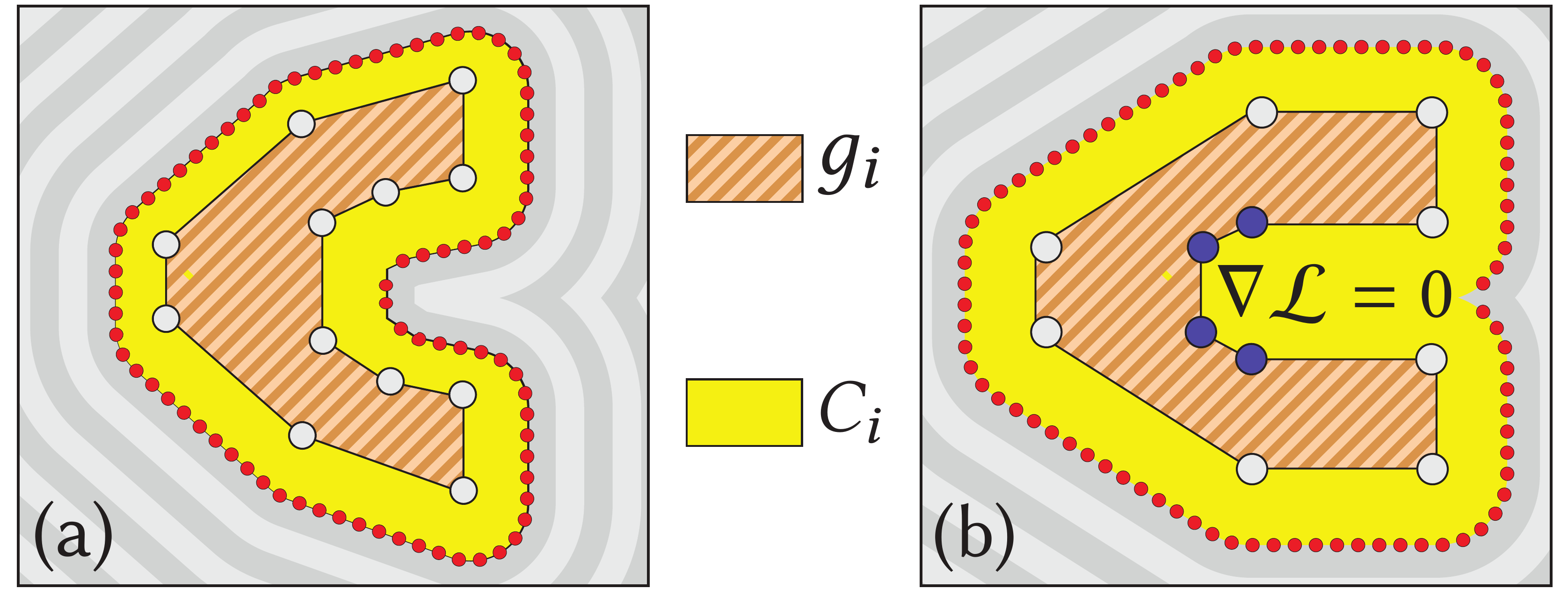}
    \caption{%
(a) We optimize the milling tool path (striped area) via the \SurfaceTermFull\ to keep the boundary of the subtracted volume (Yellow) in tight contact with the opposing part. (b) A degeneracy can arise during this process, where the \SurfaceTermFull\ has zero gradient with respect to the concave regions of the tool path (dark blue vertices).  We address the problem by introducing an additional \PathTermFull\ metric that keeps these vertices sufficiently close to the boundary.}
    \label{fig:mill_contour}
\end{figure}

While the \SurfaceTermFull~measures deviation between mating surfaces, it operates solely on the extruded boundaries \( \partial C_i \). These boundaries are produced by dilating the underlying mill paths \( g_i \), which are parameterized and optimized to restore tight coupling. This introduces a mismatch: multiple different mill paths can produce nearly identical extruded contours. As a result, optimization can become unstable, with large changes in \( g_i \) producing negligible changes in $\partial C_i$. In practice, this leads to having zero-gradients from certain parameters of \( g_i \) and consequently poor convergence in certain configurations (Figure~\ref{fig:mill_contour}). 

To address this issue, we impose an additional constraint on the mill paths.  
When a tightly coupled surface is generated by two extrusions \( (\mathcal{E}_i^a, \mathcal{E}_j^b) \) with parallel normals, their mill paths—the zero-level sets \( \partial g_i^a \) and \( \partial g_j^b \)—must stay a fixed distance apart, equal to the sum of the cutter radii.  
Each milled surface is obtained by offsetting its mill path by the cutter radius. For the two surfaces to coincide without gaps or overlap, the underlying paths must be separated by \( r_i + r_j \).

We formalize this using the \PathTermFull~(\PathTerm), which penalizes deviation from the ideal path-to-path spacing:
\begin{equation}
\label{eq:ppd}
\mathcal{M}_{\text{P}}(g^a_i, g^b_j) =
\int_{x \in \partial g^a_i} \left( \mathcal{D}(x, g^b_j) - (r_i + r_j) \right)^2 \, ds(x),
\end{equation}
where \( ds(x) \) denotes arc-length measure along the contour \( \partial g^a_i \), and \( \mathcal{D}(x, g_j^b) \) gives the signed distance from \( x \) to the opposing mill path. In practice, this loss is applied only in planar slices where both contours arise from paired extrusions with collinear axes. A symmetric version of the loss, computed over \( \partial g^b_j\) is also applied.

\subsection{Optimization}
\label{subsec:optimization}

We now detail the optimization process that transforms an artifact-free \SXGZero~program—defined under the idealized setting of zero-radius drill bits—into a physically realizable \SXGR~program that preserves tight coupling under a target drill radius \( r > 0 \).

Our optimization strategy builds on the formulation in Section~\ref{subsec:planar_obj}, where we showed that, for a single part, surface gap can be reduced to a small set of planar slice integrals—one per set of  equivalent slices. To preserve tight coupling across the full joint system, we must now perform this optimization simultaneously across all parts. 
We sample a compact set of slices that span all lateral surfaces in the coupling volume and perform co-optimization of all intersecting paths within each slice. This strategy ensures that contact is restored across all interfaces while keeping the optimization tractable. To maintain compatibility across different directions, we treat interface contours between non-aligned extrusions as fixed. This allows each slice-aligned stage to proceed independently, while preserving both millability and global coupling.
Figure~\ref{fig:planar_slices}(c-e) illustrates this process for a traditional joint: although optimization is performed on just three representative planar slices, the resulting design exhibits tight coupling across the full 3D assembly.

\paragraph{Per-Slice Optimization}
In each slice, we optimize the profile fields $\partial g_i$ of all intersecting extrusions to minimize both surface gap (\SurfaceTerm) and deviation from ideal milling path distance (\PathTerm). This requires computing losses that depend on distances between sampled points and neighboring geometry. Specifically, for each extrusion intersecting the slice, we sample points on its boundary contour \( \partial C_i \) and on its mill path $\partial g_i $, and evaluate their proximity to either other parts (for \SurfaceTerm) or paired paths (for \PathTerm).

To evaluate the \textit{Surface gap} loss \( \mathcal{L}_{\text{S}} \),  a Monte Carlo approximation of $\mathcal{M}_{\text{S}}$ (cf. equation~\ref{eq:measure_sc}),  we construct a 2D slice-wise representation of each part by projecting its material  and extrusion volumes onto the slicing plane. These are composed into a pseudo–signed  distance field using Boolean operations. While not a true Euclidean SDF, it yields  sufficiently accurate distance estimates near the contour \( \partial C_i \),  where optimization is concentrated. At each sampled point on \( \partial C_i \),  we evaluate the minimum distance to projected boundaries of other parts in the slice.
For the \textit{Milling Path Distance} loss \( \mathcal{L}_{\text{P}} \),  a Monte Carlo estimate of $\mathcal{M}_{\text{P}}$ (cf. equation~\ref{eq:ppd}), we use the exact SDFs of mill paths \( g_j \). 
At each sampled point \( x \) on the zero contour $\partial g_i$, 
we evaluate \( g_j(x) \) and penalize deviation from the target offset \( r_i + r_j \). 
This directly constrains the alignment between opposing mill paths that form 
tightly coupled surfaces.
Figure~\ref{fig:losses} illustrates how the two losses are computed: points are sampled on the surface curve to measure \( \mathcal{M}_{\text{S}} \) and on the mill paths to measure \( \mathcal{M}_{\text{P}} \).

In addition to enforcing tight coupling, we encourage the optimized geometry to remain close to the original design by using an occupancy preservation loss $\mathcal{L}_{\text{occ}}$. This term penalizes deviations from the occupancy of each part across a uniform grid of samples on the planar slice. The occupancy field is computed as a smooth function of the part's signed distance field, enabling gradient-based update.
The final optimization objective combines all three components:
\begin{equation}
    \mathcal{L}_{\text{total}} = \mathcal{L}_{\text{S}} + \lambda_{\text{P}} \cdot \mathcal{L}_{\text{P}} + \lambda_{\text{occ}} \cdot \mathcal{L}_{\text{occ}},
\end{equation}
where scalar weights \(\lambda_{\text{P}}\) and \(\lambda_{\text{occ}}\) control the relative influence of the auxiliary terms during optimization.
Further implementation details, including point sampling routines and slice-projection of expressions, are provided in the supplementary.

\begin{figure*}
    \centering
    \includegraphics[width=1.0\linewidth]{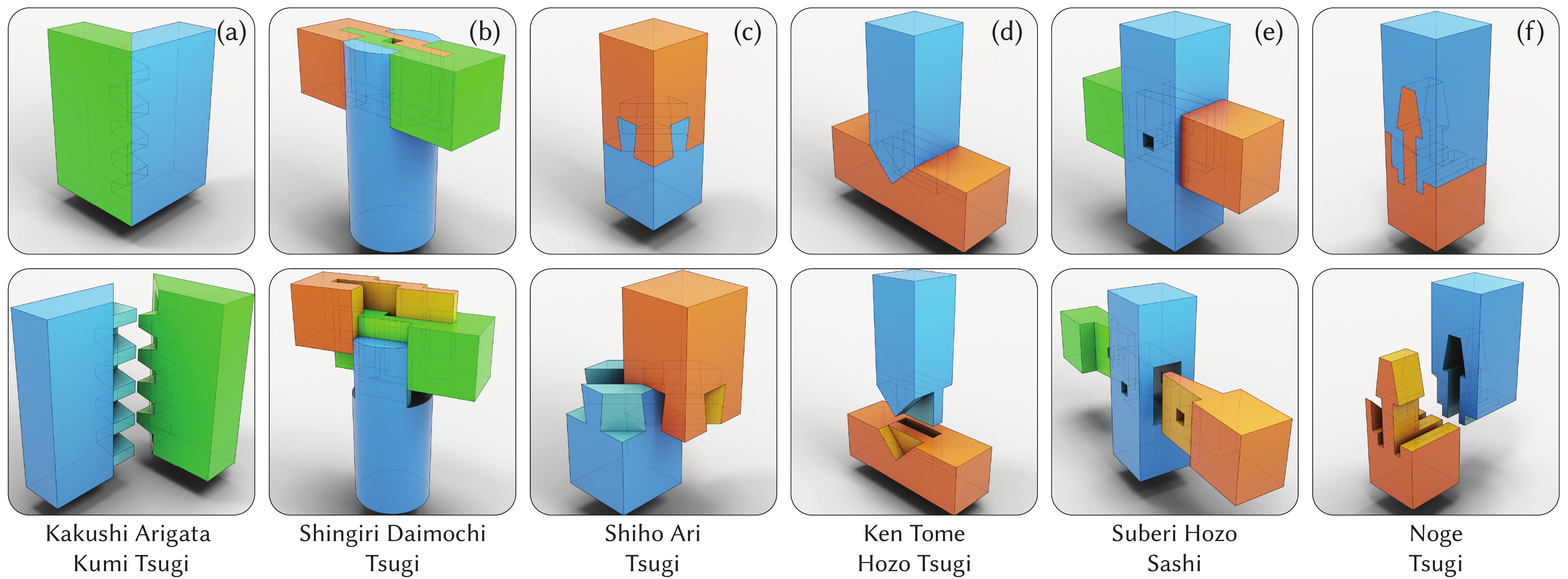}
    \caption{
    A selection of integral joints from our dataset of \DataSize~traditional designs, modeled in \SXGZero. 
Each example demonstrates a different functional or structural motif: (a) a case joint with hidden mating seams, (b)  a joint using cylindrical stock, (c) a joint with diagonal subtraction, (d) a right angled joint, (e) a joint with 3 parts, and (d) a straight connection joint. 
This dataset supports evaluation, benchmarking, and further research into CNC-fabricable joinery.
    }
    \label{fig:dataset}
\end{figure*}

\section{Evaluation}

We evaluate our approach on a curated dataset of traditional joint designs, focusing on the task of converting idealized, artifact-free programs into millable, tightly coupled assemblies. 
All experiments use a fixed drill radius of $r_d = 3.175\,$mm (1/8 inch) and assume material stock with $3 \times 3\, \text{cm}^2$ cross-section unless otherwise noted. 
We begin by describing the dataset and methods compared. We then present quantitative results that demonstrate that our method ensures tight coupling along with millability. Finally, we show fabricated outputs and design variations.

\subsection{A Dataset of Tightly-coupled Integral Joints}
\label{sec:dataset}

We construct a dataset of \DataSize~traditional joint designs using the \SXGZero~representation, based on a catalog of Japanese woodworking techniques~\cite{bracht2024japanese}. The dataset captures a broad range of integral joinery structures used in practice.
Interestingly, we found that approximately 80\% of designs in the catalog could be faithfully modeled using flat subtractive extrusions. We believe this high coverage reflects an alignment between traditional fabrication and \SXG~ representation: in both settings, all material must be removed directionally from outside the stock. More importantly, manual fabrication techniques make it difficult to construct \textit{curved} surfaces which are tightly coupled, resulting in a strong preference for planar features—precisely the type of geometry encoded by \SXG's flat extrusions.

The designs are authored using a custom visual programming tool that provides parametric control over extrusion profiles, milling directions, and depths. All parts are modeled under the zero-radius drill bit setting, yielding artifact-free geometry faithful to the source designs. Authoring the dataset required approximately~40 person-hours. 

Figure~\ref{fig:dataset} shows representative examples from the dataset. These include (a) case joints, (b) joints with cylindrical stock, (c) joints with diagonal cuts, (d) right-angled joints, (e) joints with 3 parts and (f) straight joints. 
We believe this dataset provides a concrete foundation for research in joinery modeling, fabrication-aware geometry design, and CNC-compatible procedural representations. 

\subsection{Experimental Details}

\paragraph{Comparison Methods.} 
We compare our optimization-based method against two alternatives for converting artifact-free \SXGZero~programs into millable, tightly coupled joint designs:

\begin{packed_enumerate}
\item \textit{Opening-Only (\textsc{MO}).}  
Applies morphological opening to each part’s extruded region, producing millable geometry by construction. No attempt is made to preserve surface coupling.

\item \textit{Opening \& Diff-Flip (\textsc{ODF}).}  
Begins with millable parts obtained via opening, then restores contact by applying the resulting shape differences to paired parts. That is, the volume of each part removed due to the opening operation (Diff) is directly added to the paired part (Flip). Although simple, this heuristic often yields invalid subtractions that break millability.

\end{packed_enumerate}

\paragraph{Evaluation Metrics.}
We report four metrics that collectively assess fabrication feasibility, contact quality, and geometric preservation.

\textit{Exact Millability} (\%$\mathcal{M}$) measures the percentage of designs for which all subtractions satisfy the Minkowski condition. Since morphological opening is idempotent, this is verified by checking whether each extrusion remains unchanged under opening. A design is counted as millable if every subtraction passes this test.

\textit{Coupling Success Rate} ($\mathcal{C}_\tau$) quantifies how many designs retain tight surface contact after conversion. 
For each joint, we measure the volume of overlap between parts and volume of gap i.e. space occupied by a part in the initial design that is no longer occupied by any parts.
This is done by voxelizing each part (each voxel with side length $30 / 256 = 0.11\,$mm), and counting number of voxels.
A design is considered tightly coupled if the intersection volume is within a threshold $\tau = 135\,\text{mm}^3$ (i.e. $0.5\%$ of the volume of a cube of $30\,$ mm sides). Please refer to the supplementary for evaluation over different threshold values.

To illustrate the difference between ablations of our method, we report the \textit{Median Violation Volume} ($\overline{\mathcal{V}}$), the median mismatch volume (gaps plus overlaps) across all joints.
We also measure \textit{Design Deviation} ($\overline{\mathcal{D}}$), which quantifies how much the final part geometry differs from the original artifact-free design.
We compute this by voxelizing each part before and after optimization and counting the number of non-matching voxels. 
The reported $\overline{\mathcal{D}}$ is the median of this value across all parts in the dataset.

\subsection{Quantitative Evaluation}

\begin{table}[t]
\centering
\begin{tabular}{lcc}
\toprule
\textbf{Method} & \%$\mathcal{M}$ $\uparrow$ & \%$\mathcal{C}_\tau$ $\uparrow$\\
\midrule
Opening-Only (\textsc{MO})              & 100\% &  6.25\%\\
Opening \& Diff-Flip (\textsc{ODF})      & 25\% & 96.87\%  \\
\midrule
\textbf{Ours}       & 100\% & 90.62\%  \\
\bottomrule
\end{tabular}
\caption{Comparison of methods for converting joint designs into millable, tightly coupled geometry. We report \textit{Millability} ($\mathcal{M}$) and \textit{Coupling Success Rate} ($\mathcal{C}_\tau$). Only our optimization-based method is successful at generating designs that are both millable and tightly coupled.}
\label{tab:main_results}
\end{table}

We evaluate all three methods on our dataset of \DataSize~traditional joint designs. Table~\ref{tab:main_results} summarizes the performance of each approach across millability and tight surface contact.

The \textit{Opening-only} approach achieves perfect millability ($\mathcal{M} = 100\%$), as expected from its use of morphological opening. However, it entirely fails to preserve tight coupling: almost all the designs fail to meet the contact threshold ($\mathcal{C}_\tau = 6.25\%$), and surface overlaps are visibly large.
The \textit{Opening \& Diff-Flip} method improves contact, successfully achieving tight coupling for a notable majority ($\mathcal{C}_\tau = 96.87\%$). However, 75\% of the resulting designs contain one or more subtractions that are not valid under the Minkowski condition ($\mathcal{M} = 25\%$), limiting their fabricability.
In contrast, our optimization-based approach satisfies both criteria across all designs: every joint remains fully millable by construction ($\mathcal{M} = 100\%$) while achieving tight contact on almost all mating surfaces ($\mathcal{C}_\tau = 90.625\%$). This makes it the only method to guarantee both geometric and fabrication validity at once.

\begin{figure}
    \centering
    \includegraphics[width=1.0\linewidth]{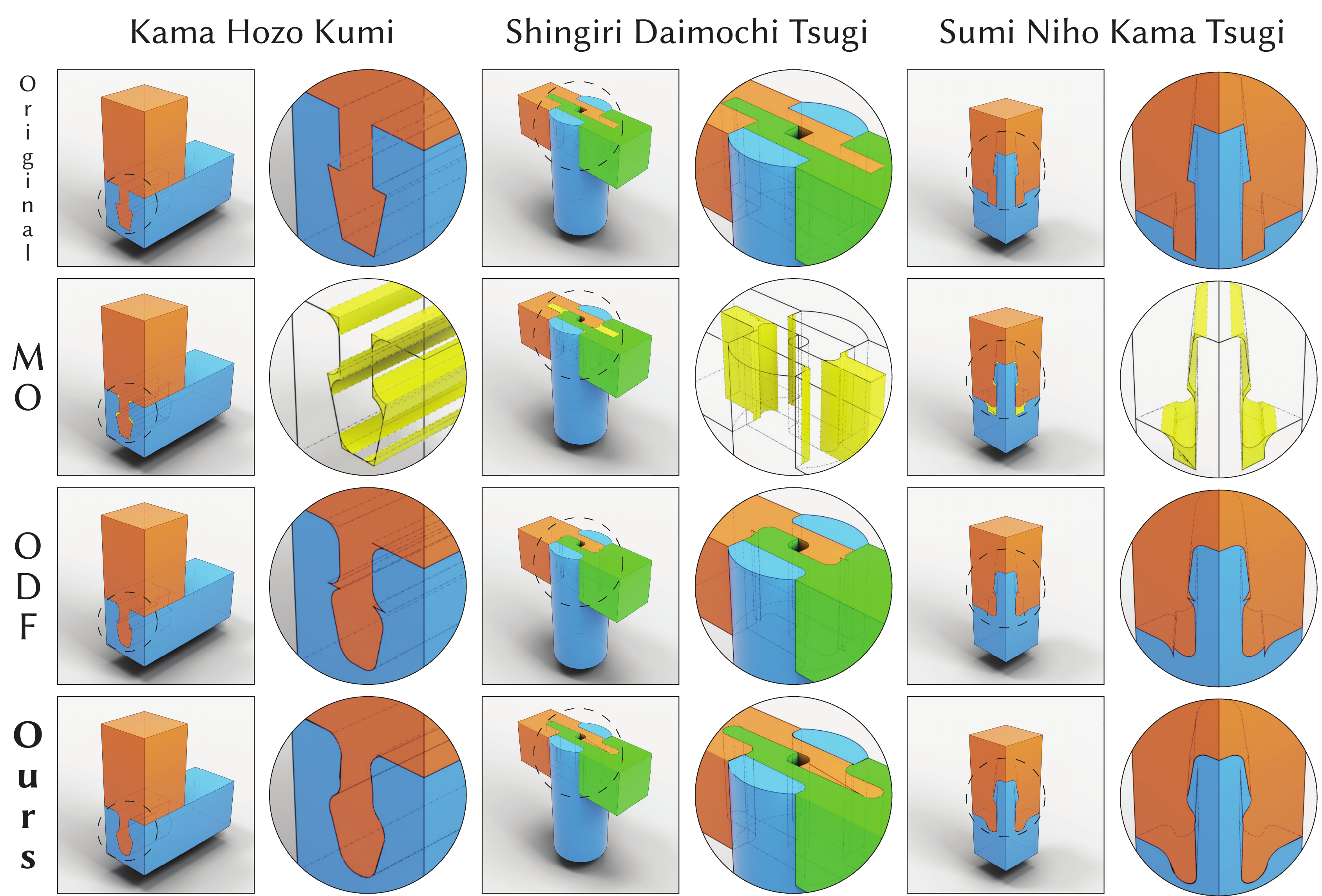}
    \caption{Comparison of methods for converting joint designs into millable, tightly coupled geometry. \textit{Opening-only} (MO)  yields overlapping parts, while \textit{Opening \& Diff-Flip} (ODF) improves contact but violates millability. Our optimization-based method achieves both tight coupling and fabrication validity. Failure regions are highlighted with zoomed insets.}
    \label{fig:baseline_fail}
\end{figure}

Figure~\ref{fig:baseline_fail} compares outputs from all three approaches across multiple joint designs. The \textit{Opening-only} (MO) method consistently produces overlapping parts that prevent assembly, while the \textit{Opening \& Diff-Flip} (ODF) variant yields better alignment but often introduces subtractions that violate millability. In contrast, our optimization-based method produces geometries that are both tightly coupled and fabricable, with minimal deviation from the original designs.

\begin{table}[t]
\centering
\begin{tabular}{lccc}
\toprule
\textbf{} & $\overline{\mathcal{V}}$ $(\text{mm}^3)$ $\downarrow$ & $\overline{\nabla \mathcal{D}}$ $(\text{mm}^3)$ $\downarrow$ \\

\midrule 
\textbf{Ours}       & \textbf{74.82 } & \textbf{456.14} \\
\midrule
No $\mathcal{L}_{\text{S}}$       & 138.23  & 483.52 \\
No $\mathcal{L}_{\text{P}}$      & 120.56  & 474.60 \\
No $\mathcal{L}_{\text{occ}}$      & 83.16  & 543.27 \\
\midrule
No gradual $r_d$ rollout             & 142.11  & 488.96 \\
No \textsc{ODF} initialization            & 238.64 & 569.59 \\
\bottomrule
\end{tabular}
\caption{Subtractive ablation of our optimization pipeline. Each row disables a single component, and we report \textit{Median Violation Volume} ($\overline{\mathcal{V}}$) and \textit{Design Deviation} ($\overline{\nabla \mathcal{D}}$). The full method yields the best performance across both metrics.}
\label{tab:ablation_results}
\end{table}
\subsection{Ablation Study}

\begin{figure*}
    \centering
    \includegraphics[width=1.0\linewidth]{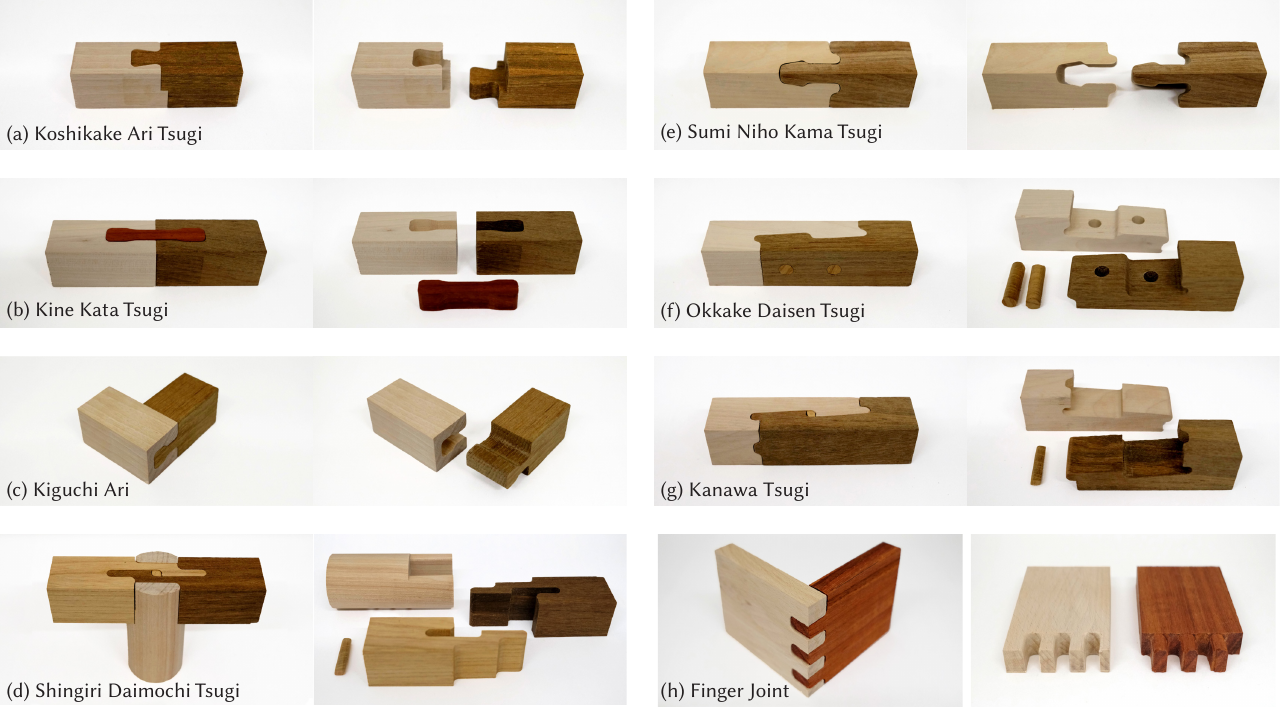}
    \caption{\textbf{CNC-Milled Physical Outputs.}
    Assembled and disassembled states of joints physically fabricated using a 3-axis CNC milling machine with a quarter-inch flat-end bit. The parts of joint (e) were positioned at a 45 degree angle during milling. The holes in joint (f) were fabricated after assembling the two pain parts of the joint, ensuring high precision despite repositioning. The parts of joint (g) were repositioned twice to accommodate the multiple milling directions. }
    \label{fig:fabricated_results}
\end{figure*}

To attribute the contribution of different components in our optimization pipeline, we conduct an ablation study by selectively removing loss terms and heuristics from our method. 
Table~\ref{tab:ablation_results} reports the effect of each modification using two metrics: average surface gap ($\overline{\mathcal{S}}$),  and design deviation ($\overline{\mathcal{D}}$).

We begin by ablating the loss functions introduced in Section~\ref{sec:tight_coupling}. Removing the \textit{Milling Path Distance} loss $\mathcal{L}_{\text{P}}$ or the \textit{Surface Gap} loss $\mathcal{L}_{\text{S}}$  leads to notable increases in surface gap—highlighting the importance of using both to achieve tight coupling. 
Omitting the \textit{Occupancy Preservation} loss $\mathcal{L}_{\text{occ}}$ increases the average design deviation ($\overline{\mathcal{D}}$), indicating that it plays a key role in preserving geometric fidelity to the original design.

We also evaluate two implementation strategies that improve optimization quality and convergence. 
First, instead of optimizing directly at the target drill radius $r_d$, we adopt a gradual rollout schedule: starting from radius zero, we increase $r$ in small increments, re-optimizing at each step. This improves stability, especially for large-radius artifacts.
Second, we initialize the profile fields $g_i$ using the output of an \textit{Opening \& Diff-Flip} (ODF) pass at radius $r_d/2$. This initialization expands the expressive range of $g_i$ early on, enabling better adaptation during optimization. Without it, we observe stagnation due to insufficient parameter flexibility. 
While adaptive reparameterization is a possible alternative, ODF initialization provides a lightweight and effective solution. As shown in Table~\ref{tab:ablation_results}, removing any of these strategies results in a significant loss in performance, illustrating that they play a key role in ensuring that the optimization is successful.

Finally, we report that our optimization process takes roughly 5 minutes per planar slice. Most joints yield within 1 to 3 slices, depending on the number of parts and the orientation of milling directions. Full optimization typically completes within $\sim$10 minutes.

\section{Fabrication and Design Outcomes}

\paragraph{Physical Fabrication}
To validate that our optimized joint designs are physically realizable, we fabricated eight joints using a 3-axis CNC machine equipped with a quarter-inch flat-end bit. Each joint was generated from an \SXG~program and optimized using our method to ensure both millability and tight coupling under the target bit radius. All square parts were cut from 3 cm wooden stock, and the cylindrical part diameter was 4 cm. The joints were assembled without glue, nails, or fasteners. Refer to the supplementary materials for further details.

Figure~\ref{fig:fabricated_results} shows the joints in assembled and dissassembled states. Each achieves precise alignment and firm contact between mating surfaces, with no visible gaps or overlaps. Importantly, the observed milling artifacts (e.g., inner corner radii) match those modeled in the optimization process, confirming that our artifact-aware representation translates accurately to physical fabrication.
Each joint required between 18–25 minutes of machine time.

\begin{figure*}
    \centering
    \includegraphics[width=1.0\linewidth]{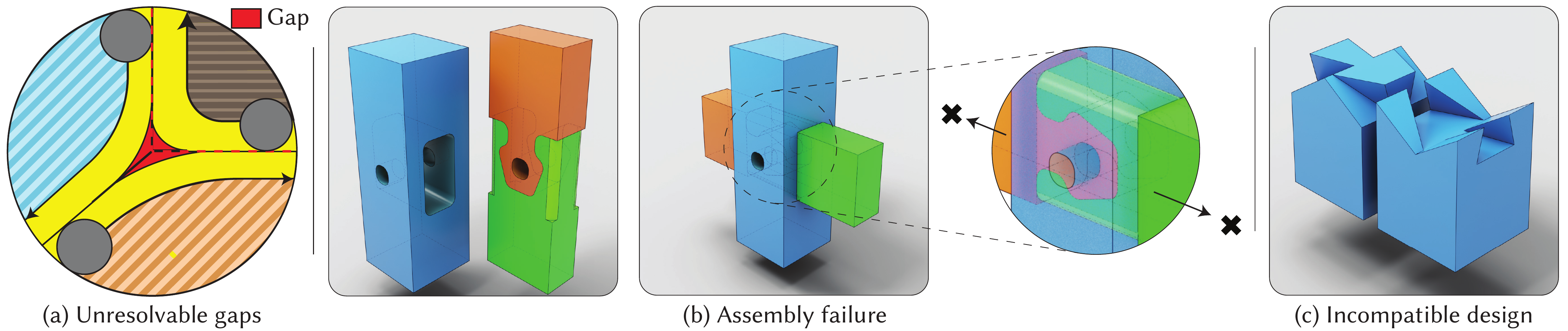}
    \caption{\textbf{Failure cases of our method} (a) Tight coupling is infeasible when three mill paths meet at a sharp corner---resulting in unavoidable gaps. (b) A joint that becomes unassemblable after optimization as \Del{the assembly sequence is not factored} \Add{our optimization does not consider assemblability as an additional requirement}. (c) The Osaka-Jo Otemon Joint, which requires subtractions incompatible with flat-end milling, and thus cannot be represented in \SXG.}
    \label{fig:failure}
\end{figure*}

\begin{figure}
    \centering
    \includegraphics[width=1.0\linewidth]{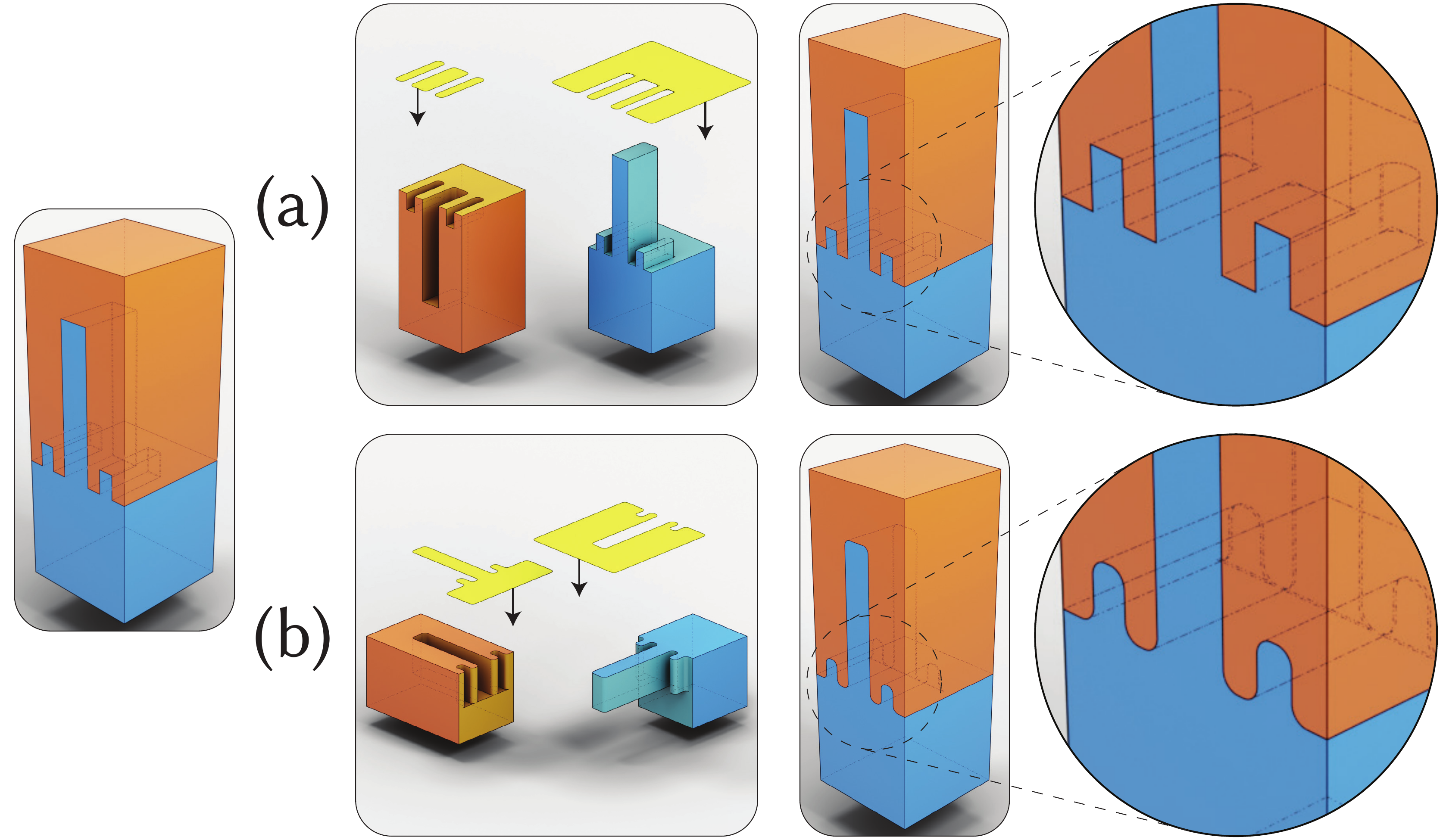}
    \caption{
    \SXG~enables controlled trade-offs between fabrication constraints and design aesthetics. Here, we show two variants of a joint which differ in milling direction. (a) Aligns all extrusions along the sliding axis, hiding artifacts internally but requiring axis-parallel milling. (b) Performs all milling laterally, improving accessibility but exposing artifacts externally. }
    \label{fig:design_variants}
\end{figure}

\paragraph{Enabling Design Exploration}
Beyond fabrication, \SXG~enables structured exploration of design alternatives. By explicitly modeling the parameters that induce milling artifacts—tool direction and drill radius—authors can evaluate how different configurations trade off aesthetic, structural, and fabrication considerations.
Figure~\ref{fig:design_variants} shows two variants of the same functional joint, each implemented via a different \SXG~program. In (a), all extrusions are aligned along the joint’s sliding axis, hiding milling artifacts inside the joint; this improves exterior appearance but requires axis-parallel milling, which can be infeasible in some setups. In contrast, (b) performs all milling laterally, ensuring accessibility on typical 3-axis machines but revealing artifacts externally.

\section{Conclusions}

In this paper, we addressed the challenge of fabricating tightly coupled integral joints with CNC milling.
CNC milling induces artifacts—such as inner corner rounding—which disrupt tight coupling, resulting in poor fit or failed assembly. 
We addressed this challenge through a two-part solution: (1) modeling how milling alters geometry, and (2) optimizing part designs to restore tight coupling despite these deviations.

To enable controllable modeling of milling artifacts, we introduced \SXGFull~(\SXG), a representation in which parts are constructed from subtractive milling operations performed with flat-end drill bits. Each operation is parameterized by a tool direction and drill radius, not only ensuring fabricability by construction but also making the source of milling artifact explicit.
To preserve tight coupling, we formalized two losses: \SurfaceTermFull, which measures geometric separation between mating surfaces, and \PathTermFull, which constrains the toolpaths that generate them. 
For geometry expressed in \SXG, both losses reduce to 1D contour integrals on planar slices, enabling tractable and accurate optimization. We optimized all extrusion profiles jointly, yielding millable geometries that maintain tight coupling under realistic fabrication constraints.

We evaluated our method on a curated dataset of \DataSize~traditional joints and found that it outperformed baseline approaches for preserving tight coupling despite milling-induced artifacts. We also fabricated 8 joints on a 3-axis CNC machine, verifying that our designs translate to physical assemblies.
By making tightly coupled integral joint design directly millable with CNC machines, our approach makes such joints more widely accessible and lays the groundwork for a new class of joinery designs.

\subsection{Limitations \& Future Work}

While our approach enables the fabrication of a broad class of traditional joints, several limitations remain—each pointing to promising future directions.

A key limitation of our approach arises when multiple concave subtraction are under tight coupling. When multiple extrusions intersect at sharp internal angles, it becomes geometrically infeasible to maintain perfect surface contact while satisfying millability constraints—resulting in small but unavoidable clearance gaps (Figure~\ref{fig:failure}(a)). This issue also arises when two extrusions interface with a fixed boundary (from non-aligned extrusions)\Add{, as in the case of the joint in Figure~\ref{fig:planar_slices}}. Addressing such cases may require hybrid fabrication strategies that go beyond 3-axis milling, such as introducing auxiliary planar cuts or incorporating secondary tools like chisels or saw blades. 
Such extensions would also be necessary to support the $\sim\!20\%$ of joints in our source catalog~\cite{bracht2024japanese} that cannot be modeled using flat-end extrusions alone. Figure~\ref{fig:failure}(b) shows the Osaka-Jo Otemon Joint, a traditional design that includes subtractions which cannot be decomposed into externally accessible flat-end milling operations.

A second limitation is the need to manually author \SXGZero~programs. While our visual programming tool streamlines this process, creating designs from scratch still requires expertise and time. Future work could explore automatic inference of subtractive extrusion programs from mesh-based geometry, allowing users to model joints in conventional CAD tools while benefiting from our representation. Additionally, improved authoring interfaces—such as motif libraries, sketch-based editing, or learning-based retrieval of common joint structures—could make \SXG\ more accessible to novice users and support faster design iteration.

Third, our system does not currently account for assembly sequencing. Although the optimized joints are tightly coupled, they may become unassemblable due to the modified geometry. In our dataset, 1 of the ~\DataSize~joints cannot be manually assembly after optimization. We show an example in Figure~\ref{fig:failure} (b). Future work could incorporate directional blocking analysis directly into the optimization loop, or constrain extrusion paths to preserve known assembly sequences.

While our system enables the fabrication of integral joints via CNC milling, the full repertoire of techniques used by master carpenters remains beyond its scope. Traditional joints often incorporate subtle construction strategies—such as intentional minuscule misalignments, or in-driven wedges—that enhance strength or aid assembly. Modeling these expert techniques and making them accessible through modern fabrication tools remains an open challenge. We view this study as a concrete step toward that broader goal.

\bibliographystyle{ACM-Reference-Format}
\bibliography{main}

\appendix

\end{document}


\title{\textsc{MiGumi}: Making Tightly Coupled Integral Joints Millable \\ --- Supplemental ---}

\author{Aditya Ganeshan}
\email{adityaganeshan@gmail.com}
\orcid{0000-0001-8615-741X} 
\affiliation{%
  \institution{Brown University}
  \city{Providence}
  \country{United States of America}
}

\author{Kurt Fleischer}
\email{kurt@pixar.com}
\orcid{0009-0007-1768-4591}
\affiliation{%
  \institution{Pixar Animation Studios}
  \city{San Francisco}
  \country{United States of America}
}

\author{Wenzel Jakob}
\email{wenzel.jakob@epfl.ch}
\orcid{0000-0002-6090-1121}
\affiliation{%
  \institution{École Polytechnique Fédérale de Lausanne (EPFL)}
  \city{Lausanne}
  \country{Switzerland}
}

\author{Ariel Shamir}
\email{arik@runi.ac.il}
\orcid{0000-0001-7082-7845}
\affiliation{%
  \institution{Reichman University}
  \city{Herzliya}
  \country{Israel}
}

\author{Daniel Ritchie}
\email{daniel_ritchie@brown.edu}
\orcid{0000-0002-8253-0069}
\affiliation{%
  \institution{Brown University}
  \city{Providence}
  \country{United States of America}
}

\author{Takeo Igarashi}
\email{takeo.igarashi@gmail.com}
\orcid{0000-0002-5495-6441}
\affiliation{%
  \institution{University of Tokyo}
  \city{Tokyo}
  \country{Japan}
}

\author{Maria Larsson}
\email{ma.ka.larsson@gmail.com}
\orcid{0000-0002-4375-473X}
\affiliation{%
  \institution{University of Tokyo}
  \city{Tokyo}
  \country{Japan}
}
\newcommand{\SXG}{\textsc{MXG}}
\newcommand{\SXGZero}{$\textsc{MXG}_{0}$}
\newcommand{\SXGR}{$\textsc{MXG}_{r}$}
\newcommand{\SXGFull}{\textsc{Millable Extrusion Geometry}}

\newcommand{\SurfaceTermFull}{\textsc{Surface Gap}\xspace}
\newcommand{\PathTermFull}{\textsc{Milling Path Distance}\xspace}
\newcommand{\SurfaceTerm}{$\mathcal{M}_{\text{S}}$}
\newcommand{\PathTerm}{$\mathcal{M}_{\text{P}}$}

\newcommand{\DataSize}{30}

\newcommand{\JointNameFull}{\textsc{Millable Kigumi}}
\newcommand{\JointName}{\textsc{MiGumi}}

\newcommand{\EFFull}{\emph{Millable Extrusion Field}}
\newcommand{\EF}{$\mathcal{E}$}

\newcommand{\aditya}[1]{\textcolor{blue}{[ADITYA: #1]}}
\newcommand{\dr}[1]{\textcolor{red}{[DANIEL: #1]}}
\newcommand{\maria}[1]{\textcolor{magenta}{[MARIA: #1]}}
\newcommand{\as}[1]{\textcolor{orange}{[ARIK: #1]}}

\newenvironment{packed_itemize}
{\begin{itemize}
    \vspace{-\topsep}
    \setlength{\itemsep}{1pt}
    \setlength{\parskip}{0pt}
    \setlength{\parsep}{0pt}
}{\end{itemize}}

\newenvironment{packed_enumerate}
{\begin{enumerate}
    \vspace{-\topsep}
    \setlength{\itemsep}{1pt}
    \setlength{\parskip}{0pt}
    \setlength{\parsep}{0pt}
}{\end{enumerate}}

\newtoggle{showchanges}
\togglefalse{showchanges}
\DeclareRobustCommand{\Add}[1]{%
  \iftoggle{showchanges}{\textcolor{blue}{#1}}{#1}%
}
\DeclareRobustCommand{\Del}[1]{%
  \iftoggle{showchanges}{\textcolor{red}{#1}}{}%
}
\DeclareRobustCommand{\Repl}[2]{
  \iftoggle{showchanges}{\Del{#2}\,\Add{#1}}{#1}%
}

\DeclareRobustCommand{\AddM}[1]{%
  \iftoggle{showchanges}{\color{blue}{#1}}{#1}%
}
\DeclareRobustCommand{\DelM}[1]{%
  \iftoggle{showchanges}{\begingroup\color{red}\cancel{#1}\endgroup}{}%
}
\DeclareRobustCommand{\ReplM}[2]{
  \iftoggle{showchanges}{\DelM{#2}\,\AddM{#1}}{#1}%
}

\DeclareRobustCommand{\Shep}[1]{%
  \iftoggle{showchanges}{\textcolor{magenta}{[Shepherd: #1]}}{}%
}


\maketitle

\begin{figure*}
    \centering
    \includegraphics[width=1.0\linewidth]{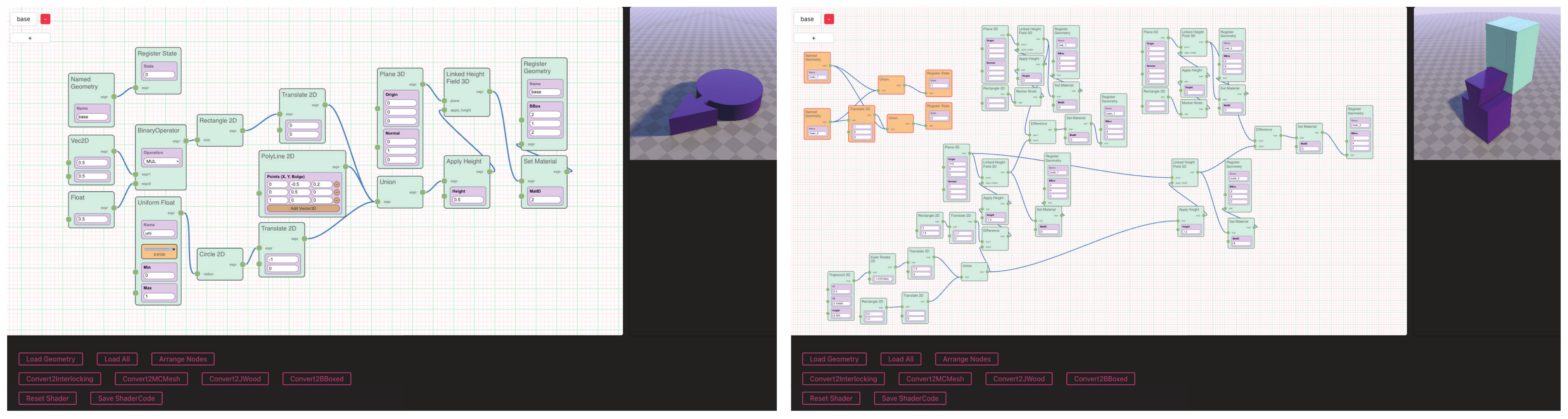}
    \caption{
    Our custom in-house node-based visual programming interface. On the left we show a simple program to create a shape. On the right we show the program to create a simple joint "Ari Tsugi". Notice the reuse of sub-graphs for 2D CSG expression between the two parts. Further, we also record the assembly sequence using a state machine. We highlight the nodes which correspond to the Assembly states in the right image (top left of the right image).
    }
    \label{fig:interface}
\end{figure*}

\section{Details: Modeling Millable Geometry}

\subsection{Cylinder Minkowski Sum Proposition}

Our representation, \SXG~leverages flat extrusions that satisfy the minkowski condition to ensure that each subtraction $V_i$ is millable. Here, we first provide the proposition and its proof that this representation is built on. 

We begin by characterizing the class of solids that can be constructed via subtractive milling\footnote{We assume all solids are regular closed subsets of $\mathbb{R}^3$, i.e., equal to the closure of their interior.}. A solid $P \subset \mathbb{R}^3$ is millable if it can be expressed as:
\begin{equation}
\label{eq:part_expr}
P = M - \bigcup_i V_i,
\end{equation}
where $M$ is the material stock and each $V_i$ is a \emph{millable} volume removed by the $i$-th milling operation.
To be millable, each $V_i$ must be a Minkowski sum\footnote{Given two sets $A, B \subset \mathbb{R}^n$, their Minkowski sum is defined as $A \oplus B = \{ a + b \mid a \in A,\, b \in B \}$.} with respect to the milling tool's rotational swept volume. The rotational sweep of flat-end uniform-radius drill-bit yields a cylinder. Therefore, this can be written as:
\begin{equation}
\label{eq:mink-3d}
V_i = X_i \oplus \text{Cyl}_i,
\end{equation}

where $X_i$ is an arbitrary shape and $\text{Cyl}_i \subset \mathbb{R}^3$ is the rotational sweep of the drill bit for the $i$-th operation. 
To enforce accessibility, $\text{Cyl}_i$ is modeled as a semi-infinite cylinder extending away from the milling direction, ensuring that each $V_i$ is accessible from outside.
This formulation guarantees millability by construction: each subtraction aligns with tool geometry and access constraints, making the resulting solid $P$ physically achievable through sequential milling.

The following proposition characterizes when a solid admits such a representation:
\begin{proposition}
\label{prop:cyl}
Let $V \subset \mathbb{R}^3$ be a solid, and let $Cyl_r$ denote the semi-infinite cylinder of radius $r$ along the $z$-axis. Then $V = X \oplus Cyl^r$ for some solid $X \subset \mathbb{R}^3$ if and only if:
\begin{enumerate}
    \item (\textbf{Minkowski Condition}) For every $z$, the horizontal slice
    \[
    C(z) = \left\{ (x, y) \in \mathbb{R}^2 \;\middle|\; (x, y, z) \in V \right\}
    \]
    satisfies $C(z) = X(z) \oplus B_r$ for some $X(z) \subset \mathbb{R}^2$, where $B_r$ is a circle of radius $r$.
    \item (\textbf{Nesting Condition}) For all $z_1 < z_2$,  $C(z_1) \subseteq C(z_2)$.
\end{enumerate}
\end{proposition}
Intuitively, this proposition states that a solid is millable with a flat-end bit of radius $r$ along a direction (e.g., the $z$-axis) if, when sliced perpendicular to that direction, each slice is a Minkowski sum of some base shape with a disk of radius $r$, and the slices nest monotonically—i.e., the volume forms a heightfield from the milling direction. Although stated here for the $z$-axis, the same conditions apply for any milling direction $\mathbf{n} \in \mathbb{S}^2$ after rotation.

\begin{proof}[Proof (if direction)]
\textbf{(\(\Leftarrow\))} Suppose \( V \subset \mathbb{R}^3 \) satisfies the following conditions:
\begin{packed_enumerate}
    \item For each height \( z \), the horizontal cross-section
    \[
    C(z) = \{ (x, y) \in \mathbb{R}^2 \mid (x, y, z) \in V \}
    \]
    satisfies \( C(z) = X(z) \oplus B_r \) for some planar shape \( X(z) \subset \mathbb{R}^2 \).
    
    \item The slices are nested: for all \( z_1 < z_2 \), \( C(z_1) \subseteq C(z_2) \).
\end{packed_enumerate}

We define the solid: $X = \{ (x, y, z) \in \mathbb{R}^3 \mid (x, y) \in X(z) \}$,
and let \( \text{Cyl}_r = B_r \times [0, \infty) \) denote the semi-infinite  vertical cylinder of radius \( r \).

\emph{We will show that} \( V = X \oplus \text{Cyl}_r \).

\noindent\textbf{(i) \( V \subseteq X \oplus \text{Cyl}_r \):}  
Let \( (x, y, z) \in V \). Then by definition, \( (x, y) \in C(z) = X(z) \oplus B_r \).  
So there exist \( (x', y') \in X(z) \), \( (u, v) \in B_r \) such that \( (x, y) = (x' + u, y' + v) \).  
Letting \( w = 0 \), we have:
\[
(x, y, z) = (x', y', z) + (u, v, w),
\]
where \( (x', y', z) \in X \) and \( (u, v, w) \in \text{Cyl}_r \).  
Thus, \( (x, y, z) \in X \oplus \text{Cyl}_r \).

\vspace{0.5em}
\noindent\textbf{(ii) \( X \oplus \text{Cyl}_r \subseteq V \):}  
Let \( (x, y, z) \in X \oplus \text{Cyl}_r \). Then there exist \( (x', y', z') \in X \), \( (u, v, w) \in \text{Cyl}_r \) such that:
\[
(x, y, z) = (x' + u, y' + v, z' + w),
\]
with $(x', y') \in X(z'), \ (u, v) \in B_r, \ w \ge 0$.
By the nesting condition, \( X(z') \subseteq X(z' + w) \). So \( (x', y') \in X(z) \), and hence \( (x, y) \in X(z) \oplus B_r = C(z) \).  
Thus, \( (x, y, z) \in V \).

\noindent\textbf{Conclusion:}  
We have shown that \( V \subseteq X \oplus \text{Cyl}_r \) and \( X \oplus \text{Cyl}_r \subseteq V \), hence:
\[
V = X \oplus \text{Cyl}_r. \qed
\]
\end{proof}

\begin{proof}[Proof (only if direction)] \textbf{(\(\Rightarrow\))}
Assume \( V = X \oplus \text{Cyl}_r \), where \( \text{Cyl}_r = B_r \times [0, \infty) \) is a semi-infinite cylinder aligned with the \( z \)-axis.

We will show that the horizontal slices of \( V \) satisfy both the Minkowski and nesting conditions stated in the proposition.

\paragraph{(i) Minkowski Condition.}
Fix any height \( z \in \mathbb{R} \), and define the horizontal slice of \( V \) at height \( z \) as
\[
S(z) = \{ (x, y) \in \mathbb{R}^2 \mid (x, y, z) \in V \}.
\]
Since \( V = X \oplus \text{Cyl}_r \), for any \( (x, y, z) \in V \), there exists \( (x', y', z') \in X \) and \( (u, v, w) \in B_r \times [0, \infty) \) such that
\[
(x, y, z) = (x', y', z') + (u, v, w).
\]
This implies \( z = z' + w \) with \( w \ge 0 \), so \( z' \le z \). Let us define
\[
X(z) = \{ (x', y') \in \mathbb{R}^2 \mid \exists z' \le z \text{ such that } (x', y', z') \in X \}.
\]
Then for any \( (x, y) \in S(z) \), we have \( (x, y) \in X(z) \oplus B_r \). Conversely, for any \( (x', y') \in X(z) \) and \( (u, v) \in B_r \), the point \( (x'+u, y'+v, z) \in V \). Therefore,
\[
S(z) = X(z) \oplus B_r.
\]

\paragraph{(ii) Nesting Condition.}
Let \( z_1 < z_2 \). Then by construction, \( X(z_1) \subseteq X(z_2) \), since \( \{ z' \le z_1 \} \subseteq \{ z' \le z_2 \} \). Taking Minkowski sums with \( B_r \), we get:
\[
S(z_1) = X(z_1) \oplus B_r \subseteq X(z_2) \oplus B_r = S(z_2),
\]
as required.
\end{proof}

\begin{figure}
    \centering
    \includegraphics[width=1.0\linewidth]{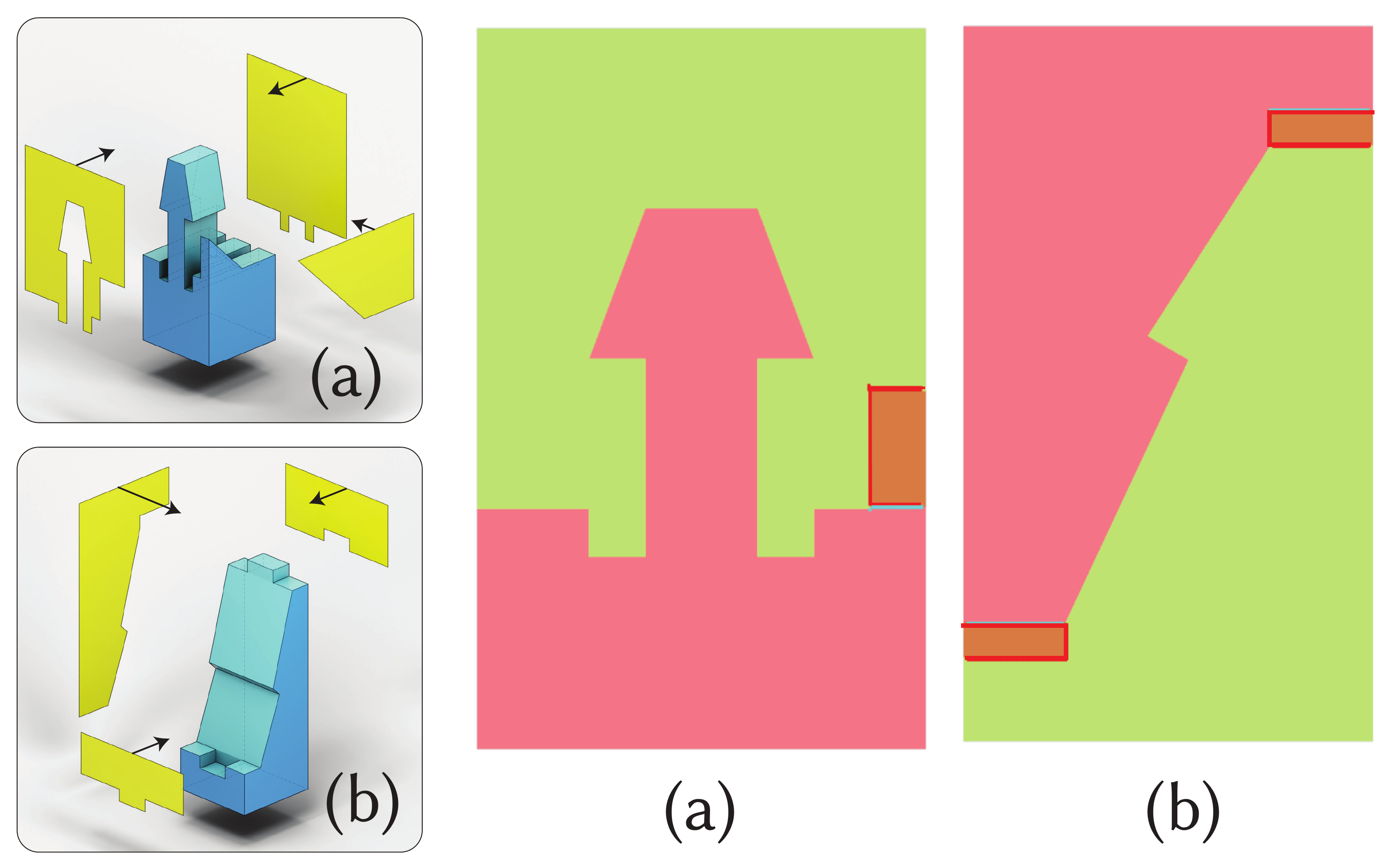}
\caption{Fixed interface constraints during slice-wise optimization. We show two example joints with the interface with non-aligned fixed extrusions highlighted in red. These fixed interfaces are preserved by enforcing a strong boundary loss, allowing the optimization to proceed on a compact set of slice-aligned contours without hampering tight coupling between non-aligned extrusions.}
    \label{fig:interface}
\end{figure}

A constructive way to satisfy these conditions is to extrude a 2D region $C_i \subset \mathbb{R}^2$, embedded in a plane orthogonal to the milling direction $\mathbf{n}_i$, over a semi-infinite interval $(-\infty, h_i)$. This ensures that the volume forms a heightfield along $\mathbf{n}_i$, satisfying the nesting condition. 
If $C_i$ additionally satisfies the Minkowski condition (i.e., $C_i = X_i \oplus B_r$ for some $X_i \in \mathbb{R}^2$), then the resulting solid is millable with a flat drill bit of radius $r$.
This leads naturally to our core primitive: the subtractive extrusion, which forms the foundation of the \SXG\ representation.

\subsection{Valid Dilation and Erosion Operations}

As described in the main paper, each extrusion field $\mathcal{E}_i$ is defined by a 2D signed distance function (SDF) $f_i$, constructed using symbolic expressions composed of primitives and Boolean operations such as union, intersection, difference, and complement. While this representation offers high expressivity, the general protocol for evaluating such expressions using min/max functions yields only pseudo-SDFs—functions that preserve the sign of the distance but do not provide accurate Euclidean values~\cite{exact_sdf_neural}.

This approximation is detrimental for morphological operations such as dilation and erosion, which rely on exact distance values to compute correct offsets. To address this, we convert each symbolic expression into a PolyCurve representation: a closed non-intersecting polygonal chain composed of straight-line segments and circular arcs. We ensure that all Boolean operations (e.g., intersections or differences) are performed such that no two primitives have intersecting boundaries. This guarantees that min/max-based SDF evaluation produce exact SDFs.

Beyond improving accuracy for morphological operations, PolyCurve conversion also facilitates optimization. The original symbolic representation may lack sufficient degrees of freedom to reduce loss terms effectively. In contrast, PolyCurves provide direct, parameterized control over geometric elements, improving convergence during optimization.

\subsection{Authoring Interface}

We construct \SXGZero~programs for each joint using a custom in-house node-based visual programming interface, shown in Figure~\ref{fig:interface}. The interface exposes symbolic construction of extrusion profiles via basic geometric primitives and 2D CSG operations. Though minimal in design, it supports parametric control and real-time feedback.

Internally, each expression is compiled into GLSL shader code and rendered using ray marching—enabling direct visualization without conversion to triangle meshes. Our renderer builds on Inigo Quilez's ShaderToy pipeline, allowing fast previews of complex joint geometries.

While the current interface is targeted at expert users, we view it as a foundation for broader tooling. Making this system usable by non-experts remains important future work.

\section{Details: Restoring Tight Coupling}

\subsection{Fixed Interface Constraint in Equivalence Slice Sets}

Our optimization process operates slice-by-slice, but many slices are redundant in their contribution to the coupling loss. To reduce computational cost, we identify a compact set of representative slices that together cover the lateral surface of the joint. We refer to this set as the \textit{equivalence slice set}. Each slice is selected from a set of planar extrusions that share tool direction and overlapping coupling regions.

However, not all interfaces are optimized simultaneously. Interfaces with extrusions from non-aligned directions—i.e., those not belonging to the current slicing direction—are treated as fixed. Figure~\ref{fig:interface} (a,b) shows examples of such non-aligned interfaces (highlighted with lines) which are held constant during a particular slice-aligned optimization stage.

To preserve surface continuity at these boundaries, we enforce a strong \textit{boundary loss} that penalizes deviation from the current surface geometry. This effectively treats the extrusions from other directions as fixed walls, guiding the optimization toward solutions that remain compatible with them.

\begin{algorithm}[h]
\caption{Optimization from \SXGZero~Program to \SXGR~Program}
\label{alg:uplift}
\begin{algorithmic}[1]
\Require \SXGZero~Program $\mathcal{P}_0$, target drill radius $r$
\For{each direction $\mathbf{n}$}
    \State Identify extrusion fields aligned with $\mathbf{n}$
    \State Sample slicing planes orthogonal to $\mathbf{n}$
    \State Group extrusions into equivalence sets $\{\mathcal{S}_1, \dots, \mathcal{S}_K\}$        
    \For{each equivalence set $\mathcal{S}_k$}
        \State Freeze misaligned boundary constraints
        \For{$t = 1$ to $T$}
            \State Increase dilation rate $r_d \leftarrow \text{schedule}(t)$
            \State Sample points on $\partial g$ \& $\partial C_i$ on each slice
            \State Compute total loss $\mathcal{L}_{\text{total}}$
            \State Update $g_i$ parameters via gradient descent
        \EndFor
    \EndFor
\EndFor
\State \Return Final \SXGR~Program $\mathcal{P}_r$
\end{algorithmic}
\end{algorithm}

\subsection{Optimization}

Algorithm~\ref{alg:uplift} summarizes the full pipeline used to optimize \SXGZero~program to its fabricable \SXGR~form.  

None of the algorithm’s steps require manual input.  
Extrusion fields aligned with a sampled direction~$\mathbf{n}$ are detected via their dot product with~$\mathbf{n}$ (step 2).  
Slices are then generated at regular intervals along this direction as intersections between the fields and planes at the sampled positions (step 3).  
Finally, to form equivalence sets, we assign each slice a signature based on the extrusions it intersects with (each slice's signature is the set of signatures of the extrusions it intersects with) and group slices that share the same signature to form the equivalence sets (step 4).

\paragraph{Sampling points on Part contours $\partial C_i$}:
Let $\mathcal{E}_i$ be an extrusion aligned with the slicing plane $z_k$, defined via a 2D signed distance function $g_i : \mathbb{R}^2 \rightarrow \mathbb{R}$. 
Its dilated contour $\partial C_i$ is defined by the $r$-level set: $\partial C_i = \{x \mid f_i(x) = r\}$.
Now, $f_i$ is constructed from boolean compositions over \textit{PolyCurve} primitives—closed curves composed of line and arc segments.  
Therefore, we can sample points $x_\ell \in \partial C_i$ in a differentiable way using a three-step approach:
\begin{packed_enumerate}
    \item \textbf{Offset curve primitives}: Sample points on each line and arc segment of the polycurve and offset them outward along the normal by distance $r$.
    \item \textbf{Handle corners}: At each vertex $v_i$ of the polycurve, generate a circular arc of radius $r$ and sample along it.
    \item \textbf{Contour filtering}: Reject any point $x$ for which $f_i(x) \neq r$, ensuring that retained samples lie on the true offset contour.
\end{packed_enumerate}
This sampling strategy avoids constructing a closed-form expression for the offset curve and remains fully differentiable with respect to the parameters of $f_i$.

\paragraph{Per-part 2D Signed Distance Field}
Given a sample $x_\ell \in \partial C_i(z_k)$, we compute its signed distance from the 2D projection of another part $P^b$ in the same slice plane. 
This part is represented as $P^b(z_k) = M{(z_k)} - \bigcup_j \overline{\mathcal{E}_j(z_k)}$.
Since all components—$M{(z_k)}$ and $\overline{\mathcal{E}_j(z_k)}$—are represented as signed distance functions (SDFs), we construct a pseudo-SDF for $P^b(z_k)$ using min/max composition. 
The distance from part $\partial P^b(z_k)$ can then be evaluated as:
$\text{SDF}_{P^b}(x_\ell)$. 
Although this pseudo-SDF may deviate from the true Euclidean SDF away from the boundary, it matches closely near the zero-contour, where our optimization is concentrated. We provide additional discussion and analysis in the supplementary.

\paragraph{Occupancy Preservation Loss}

We define occupancy as a continuous field estimated using the signed distance function $SDF_P$ for each part $P$. Let $O(x)$ be the binary occupancy of the original shape (pre-optimization), and let $\hat{O}_\theta(x)$ denote the soft occupancy field of the optimized part at point $x$, derived from its SDF using a sigmoid activation:
\begin{equation}
    \hat{O}_\theta(x) = \sigma(-\alpha \cdot \mathcal{D}_P(x)),
\end{equation}
where $\alpha$ controls the sharpness of the transition. We then define the occupancy loss as:
\begin{equation}
    \mathcal{L}_{\text{occ}} = \sum_{x \in \mathcal{G}} \left( \hat{O}_\theta(x) - O(x) \right)^2,
\end{equation}
where $\mathcal{G}$ is a uniform grid of sample points within the coupling volume. This loss encourages the optimized parts to remain close to their original geometry in both shape and extent, acting as a regularizer on the optimization.

\begin{figure}[t]
    \centering
    \includegraphics[width=\linewidth]{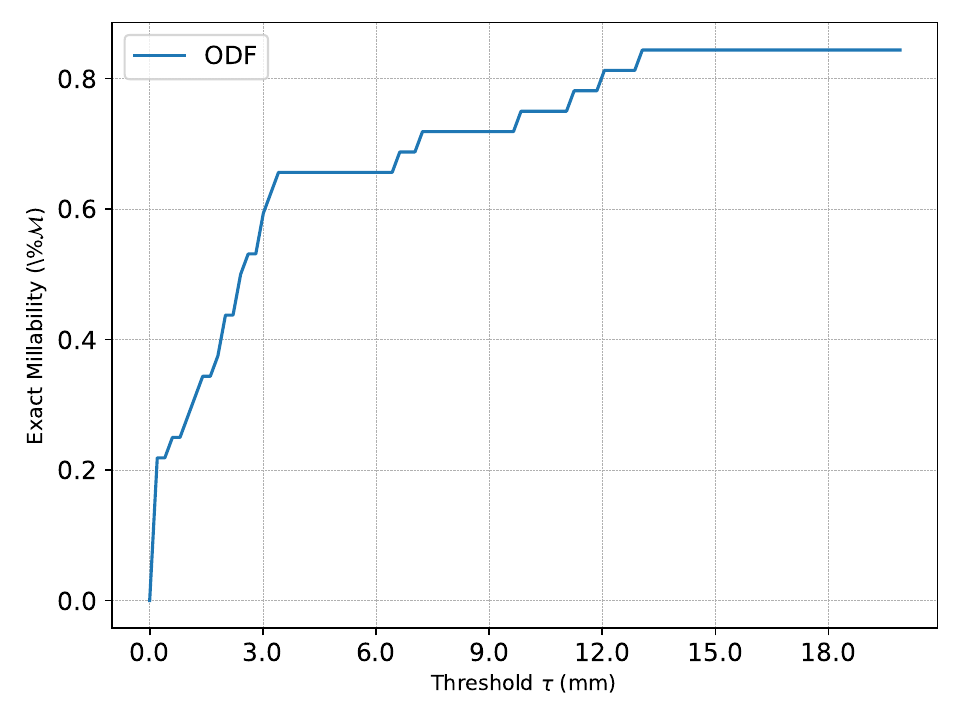}
    \caption{
    Threshold sensitivity analysis. 
    Left: \textit{Exact Millability} (\%$\mathcal{M}$) of the ODF method across varying distance thresholds. 
    Trends remain consistent across a wide range, confirming that the results in the main paper are not sensitive to specific parameter choices.
    }
    \label{fig:threshold_plot_1}
\end{figure}
\begin{figure}
    \centering
    \includegraphics[width=1.0\linewidth]{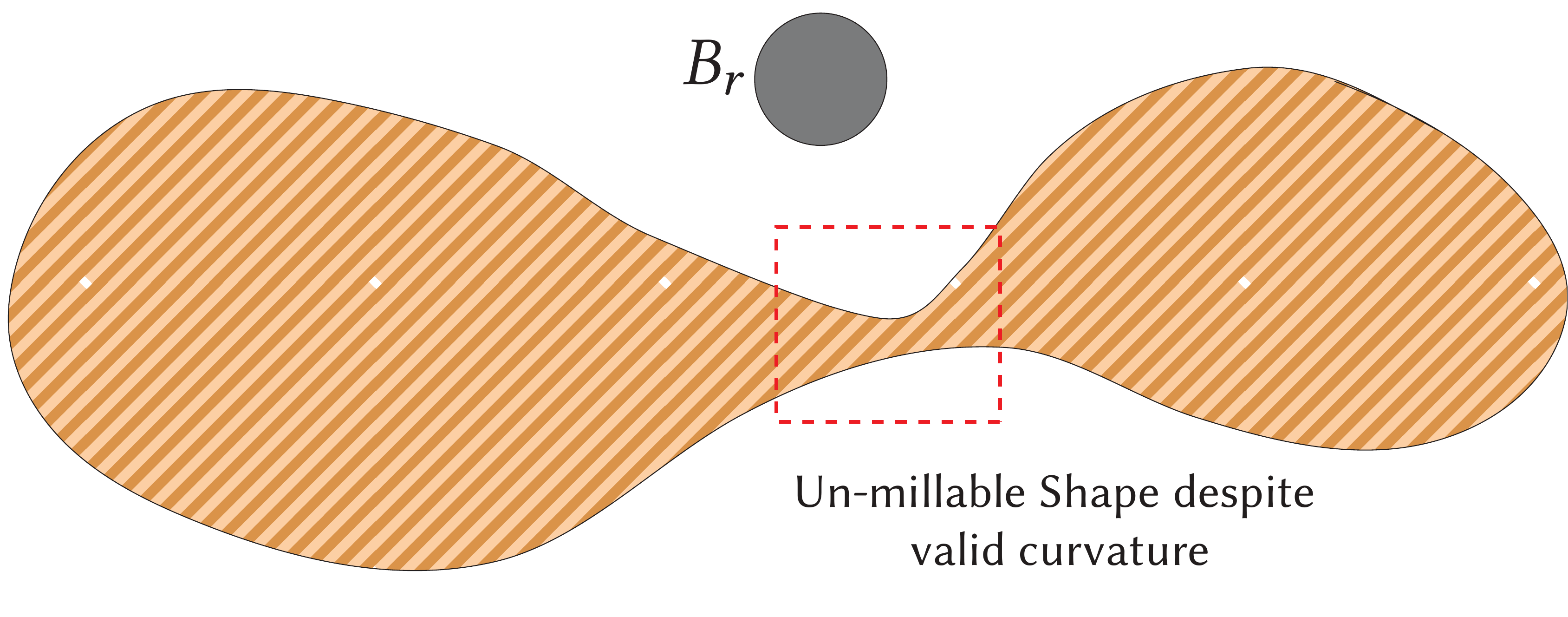}
\caption{For millability of a given shape with a drill bit of radius $r$, constraining the boundary curvature to be less than $1/r$ is insufficient.}
    \label{fig:curvature_failure}
\end{figure}

\begin{figure}[t]
    \centering
    \includegraphics[width=\linewidth]{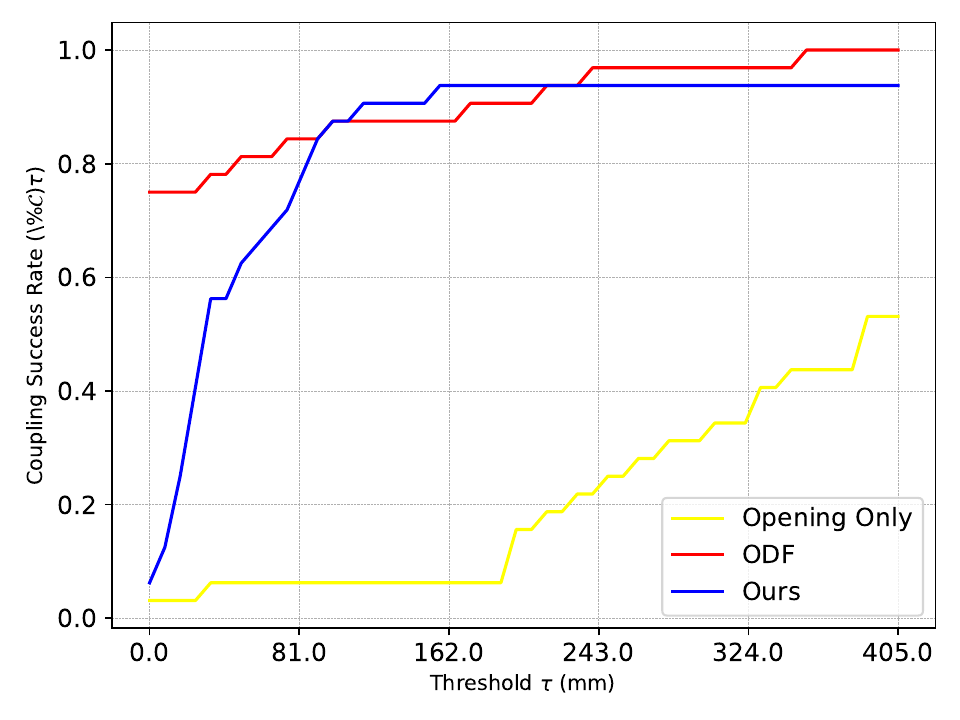}
    \caption{
    Threshold sensitivity analysis.
    Right: \textit{Coupling Success Rate} ($\mathcal{C}_\tau$) across the same thresholds. 
    Trends remain consistent across a wide range, confirming that the results in the main paper are not sensitive to specific parameter choices.
    }
    \label{fig:threshold_plot_2}
\end{figure}

\begin{figure*}
    \centering
    \includegraphics[width=0.95\linewidth]{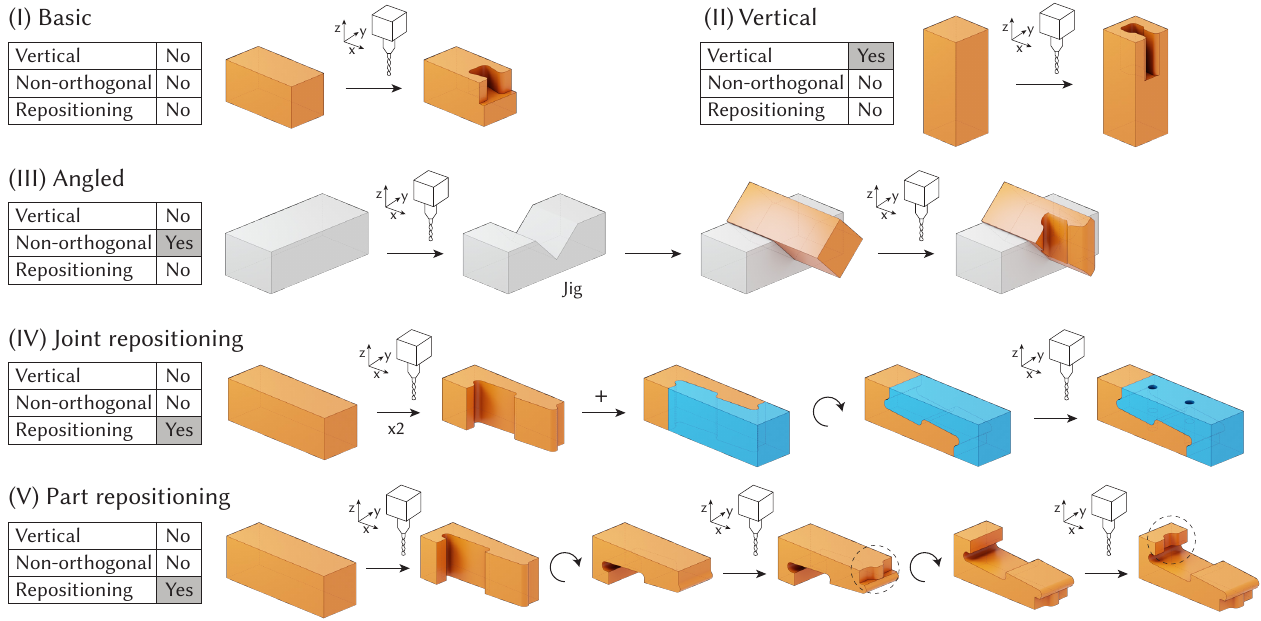}
    \caption{
    The five different types of milling procedures used to fabricate the eight physical outputs in Fig. 15 of the main paper. } \label{fig:supp_fabrication_procedure_illustration}
\end{figure*}

\subsection{Alternate Approaches}

We considered two alternate strategies for generating millable, tightly coupled geometries, but found them either insufficient or impractical in our setting.

\paragraph{Iterative ODF}
One possible approach is to alternate morphological opening with Diff-Flip updates, progressively restoring coupling after each step. However, in practice, this method encounters major hurdles. First, most 2D boolean libraries lack robust support for polycurve representations involving both lines and arcs, making repeated set operations error-prone. Even with rasterized occupancy grids and morphological operations, the procedure becomes unstable after just a few iterations. Moreover, convergence is not guaranteed, and thin features are often lost due to over-erosion. While simple in principle, this approach fails to yield consistent or controllable results.

\paragraph{Curvature Constraint.}
Another approach can be to constrain the curvature of 2D contours used in subtractive profiles. Since flat-end tools cannot produce sharp turns, one might attempt to ensure that the curvature of all profile paths remains below \(1/r\), where \(r\) is the milling tool radius. When a contour is at the interface of two joints, we can then constraint the absolute curvature to be less than \(1/r\). While this condition is necessary, it is not sufficient: a curve may have bounded curvature but still violate millability if opposing path segments come too close. Figure~\ref{fig:curvature_failure} shows a counterexample where the curvature constraint holds, yet the tool cannot pass through due to local underclearance. Millability imposes global constraints on offset distance, not just local curvature bounds.

One might attempt to strengthen the curvature condition by also enforcing a minimum local thickness, requiring that opposing segments of a contour remain at least one tool diameter apart.  
However, implementing such a rule is non-trivial: it demands reliable detection of thin regions and careful reshaping to increase clearance without altering intended contact geometry.  
Moreover, for joints involving more than two parts, it is unclear how curvature and thickness bounds should be formulated across several interacting boundaries.  
These challenges limit the practicality of a curvature-plus-thickness strategy and motivate the global offset-distance formulation adopted in our method.

\subsection{Optimization and Implementation Details}

We implement our optimization pipeline in PyTorch, using the AdamW optimizer with a learning rate of 0.003. Each slice-level optimization is run for 250 iterations, and we select the iteration with the lowest boundary loss as the final result. We ran all our experiments on an Alienware workstation equipped with a Intel i9 11900K CPU, a Geforce RTX 3090 GPU and 64 GBs of DDR4 RAM.

Prior to optimization, we convert all input extrusion profiles to non-intersecting polycurve representations. Due to occasional instability in boolean operations over polycurves (e.g., unresolved self-intersections or missing arcs), we manually corrected a small number of cases. These corrections apply only to the initial geometry and do not affect the optimization procedure. Additionally, for two joints in our dataset, we adjust the threshold used for detecting coupling boundaries due to noisy overlaps in the manually authored input geometry.

For joint assemblies involving three or more parts, we observe instability in the \PathTerm~gradient when a point on one milling path lies near multiple opposing paths. To mitigate this, we introduce a weighting scheme that downweights gradient contributions from such ambiguous regions. Specifically, we compute an entropy-based score over the top-$k$ nearest opposing points and suppress gradients where the entropy exceeds a threshold, indicating poor localization of a unique coupled path. This improves convergence and avoids incorrect updates due to competing path associations.

\section{Evaluation}

Figure~\ref{fig:threshold_plot_1} plots \textit{Exact Millability} (\%$\mathcal{M}$) for ODF as a function of the distance threshold used to verify the Minkowski condition. 
Figure~\ref{fig:threshold_plot_2} shows the corresponding \textit{Coupling Success Rate} ($\mathcal{C}_\tau$) across thresholds.
Together, these plots demonstrate that the trends reported in the main paper are robust and not the result of hand-tuned thresholds.

\section{Dataset}

Our dataset consists of \DataSize~traditional integral joint designs, each modeled parametrically using our \SXGZero~representation. These parametric programs allow for continuous variation in dimensions, proportions, and milling configurations—enabling the synthesis of a much larger family of design variants from each base joint.

We plan to expand the dataset to include additional designs from historical catalogs and contemporary applications.
Figure~\ref{fig:dataset_overview} provides an overview of all joint designs currently included in the dataset.

\section{Details of the Physical Fabrication Process}
We fabricated eight joints using a 3-axis CNC machine. 
The references (a)-(h) that follows refers to these physically fabricated joints (refer to in fig. 15 in the main paper).
Joints (a) \textit{Koshikake Ari Tsugi} and (b) \textit{Kime Kata Tsugi} were fabricated in the most straight-forward manner, without repositioning and in flat orthogonal positions (\cref{fig:supp_fabrication_procedure_illustration}-I). 
For joints (c) \textit{Kiguchi Ari} and (d) \textit{Shimigiri Daimochi Tsugi}, one part each were fabricated in a vertical orthogonal position (\cref{fig:supp_fabrication_procedure_illustration}-II).
Joint (e) \textit{Sumi Niho Kama Tsugi} was positioned at a 45 degree angle to achieve the diagonal cut. To reliably position the material at the 45 degree angle, we first milled out a jig, and then positioned material there before milling (\cref{fig:supp_fabrication_procedure_illustration}-III).
Joint (f) \textit{Okkake Daisen Tsugi} was repositioned to accommodate the milling of holes facing in a different direction from the main geometries (\cref{fig:supp_fabrication_procedure_illustration}-IV).
For this joint, we first fabricated the two main parts. Then we assembled them, and cut out the holes for the plugs in the assembled state. In this way, there is no loss in fabrication precision in terms of the tightness of the coupling despite the repositioning---even if the repositioning is slightly inexact, the holes in the two different parts will still be perfectly aligned because there are cut together.
Joint (g) \textit{Kanawa Tsugi} requires a different type of repositioning, namely, millings from three directions per part (\cref{fig:supp_fabrication_procedure_illustration}-V). This process is more error-prone, so fabrication with a 4-or-more axis CNC machine would have been preferred if we had had access to such a machine. Nonetheless, we managed to fabricate a functional joint with our 3-axis machine setup by careful repositioning.

The CNC machine we used are of the brand \textit{Shopbot} and model \textit{Desktop MAX with Aluminum T-Slot Deck}. The milling paths were created by importing meshes (*.stl files) of our system outputs to the software \textit{VCarve}, and then generating roughing tool paths with clearance set to 0.0. To create a small tolerance of the joint, we set the milling bit size to slightly (about 0.05 inches) smaller than the actual mill bit size (1/4 inch). This will result in a slight over-cut, and can the amount can be adjusted depending on the desired tightness of the joint.

\section{Additional Results}

\begin{figure}
    \centering
    \includegraphics[width=1.0\linewidth]{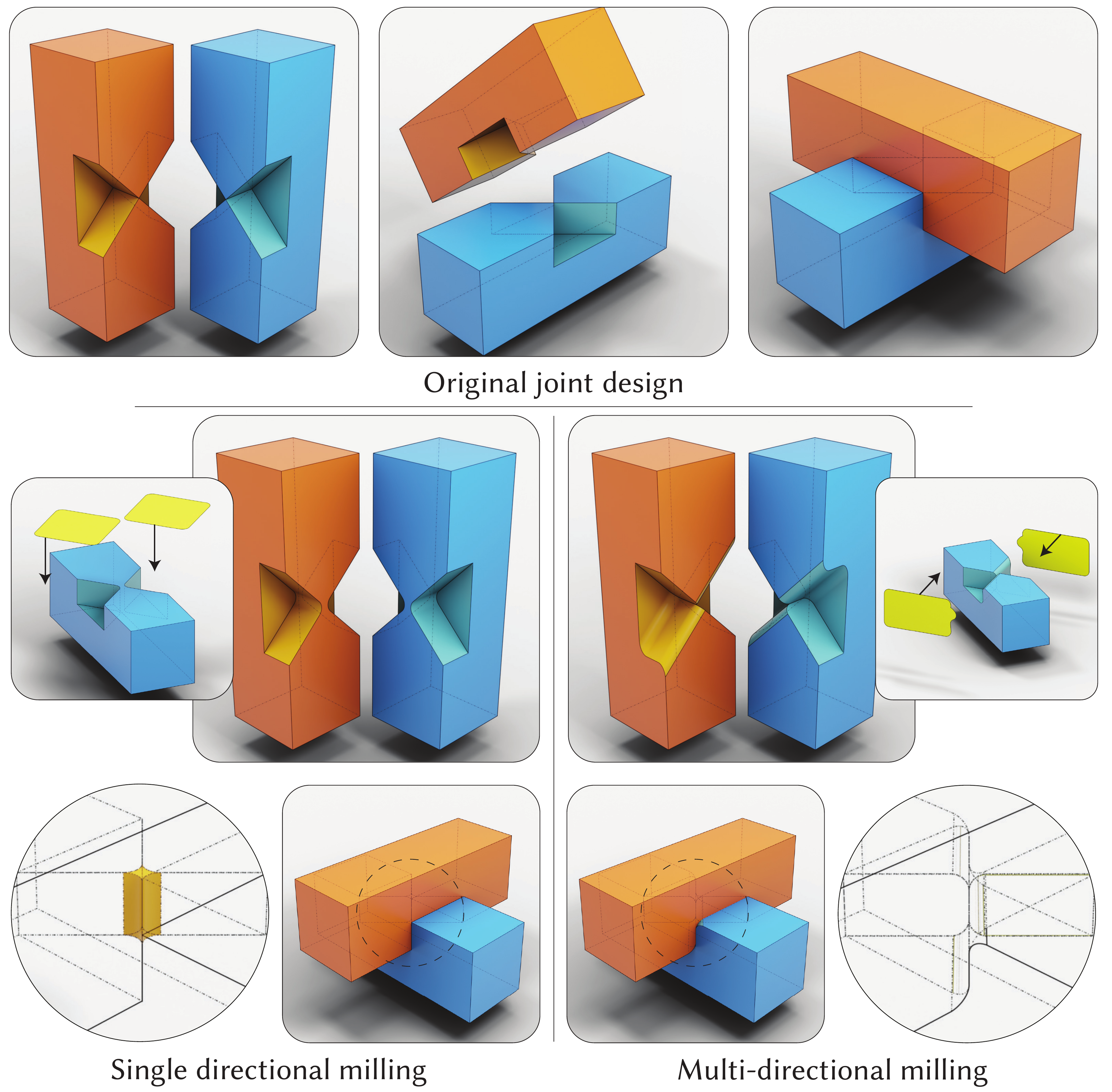}
    \caption{
    For certain joints, multi-directional milling is critical to ensure that tight coupling is feasible. While fabrication of individual parts with a single subtractive operation is feasible, it results in shapes that cannot be coupled, despite optimization. In contrast, with two oblique subtractions, we can generate a design that is millable and tightly coupled (after optimization).
    }
    \label{fig:new_designs}
\end{figure}

\begin{figure*}
    \centering
    \includegraphics[width=1.0\linewidth]{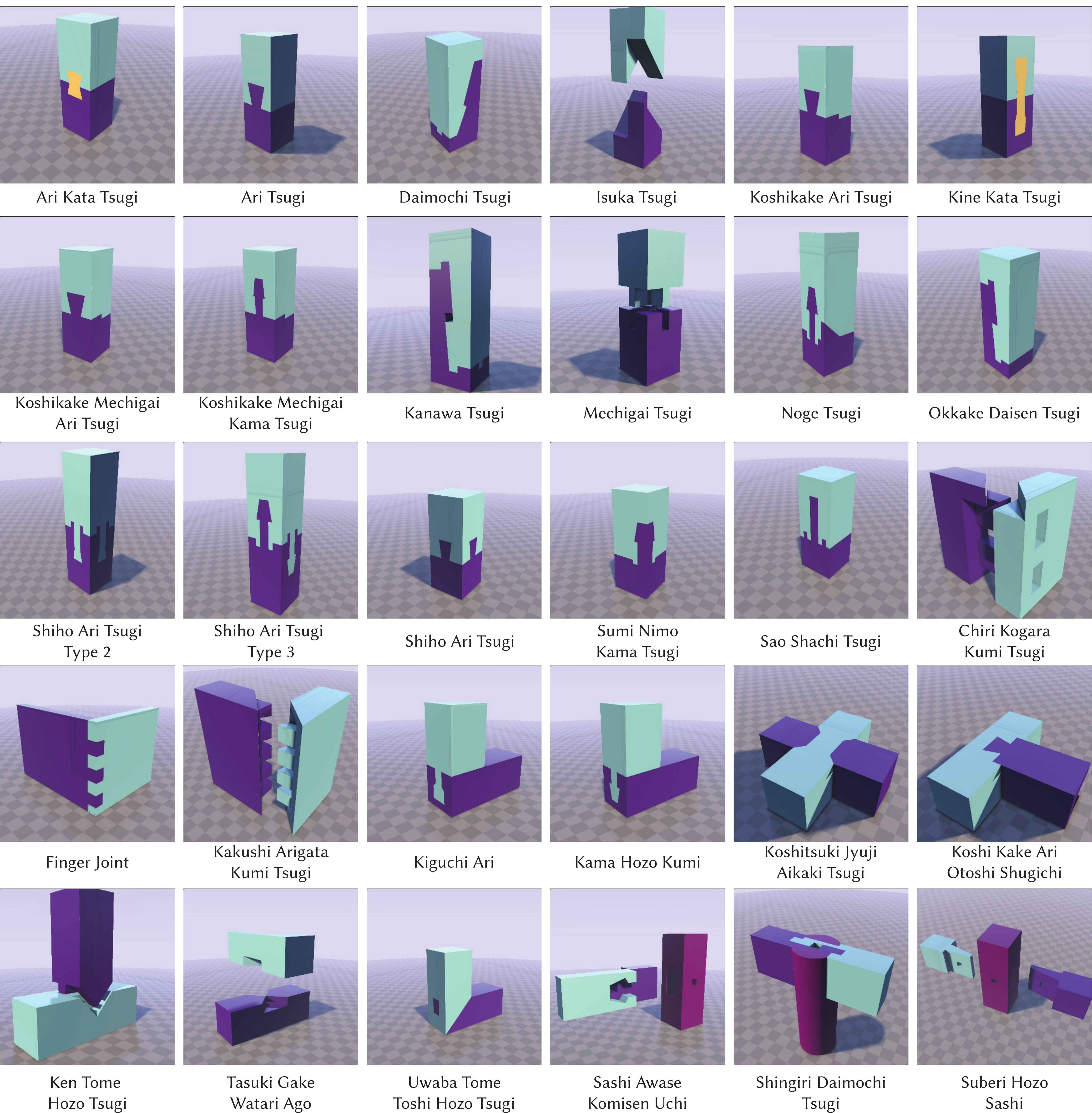}
    \caption{
    Our Full dataset of \DataSize~Joints.
This dataset supports evaluation, benchmarking, and further research into CNC-fabricable joinery.
    }
    \label{fig:dataset_overview}
\end{figure*}

\paragraph{Expanding the Design Space} 
Our method accommodates complex joint configurations that lie outside the scope of prior systems. 
The use of general CSG-like 2D expressions enable us to cover a wider spread of designs.
Furthermore, for certain joints, multi-directional milling is critical to ensure that tight coupling is feasible. Figure~\ref{fig:new_designs} shows such a joint. While fabrication of individual parts with a single subtractive operation is feasible, it results in shapes that cannot be coupled, despite optimization.
In contrast, with two oblique subtractions, we can generate a design that is millable and tightly coupled (after optimization).

\section{Frequently Asked Questions (FAQs)}
During review, several thoughtful questions were raised about our approach; we summarize them here for clarity.

\paragraph{Does the method require substantial manual modeling?}  
Some manual input is needed because no dataset of CNC-adapted traditional joints exists, making data-driven automation infeasible. Moreover, joint design involves subjective choices—such as where to allow small milling artifacts—that benefit from designer control. Our interactive editor streamlines this process by enforcing millability while preserving flexibility. Future work will explore partial automation of common edits and a user-study evaluation.

\paragraph{Is the approach limited to simple geometry or assemblies?}  
Traditional joinery was historically shaped with planar tools such as chisels and saws, and rarely involves free-form 3D surfaces or elaborate multi-stage assemblies. Our method targets this regime and does not attempt to cover joints requiring sculpted surfaces or complex assembly choreography.

\paragraph{In Fig.~4(c), why are some joint boundaries rounded while others remain sharp?}  
We apply a morphological opening with radius \(r\) to the subtraction region. This guarantees millability but differs from uniform filleting: only concave corners inside the removal region (yellow) are rounded, while other edges remain unchanged.

\paragraph{Is the method specific to 3-axis CNC milling?}  
The geometry produced by our algorithm is compatible with 4- and 5-axis mills. We emphasise 3-axis setups because they are widely available in standard woodworking shops, while higher-axis machines can execute the same extrusions with fewer reorientations.

\paragraph{Why assume a single cutter radius? Could multiple tools improve efficiency?}  
Our formulation adapts geometry for the smallest finishing radius, ensuring tight coupling and clearance. Standard practice—roughing with larger cutters and finishing with a small tool—is fully compatible with this geometry and supported by existing CAM software; our focus is on geometric correctness, not minimising machining time.

\paragraph{Does the approach guarantee global accessibility for the cutter?}  
Every subtraction is constructed as a semi-infinite extrusion along a fixed direction (Supplementary Section~1), which guarantees a straight-line tool path along that direction, assuming feasible fixturing.

\paragraph{How are 2D mill path contours represented?}  
They are stored as a sequence of straight segments and circular arcs, using a vertex–bulge format. For implementation details, please refer to the released code.

\bibliographystyle{ACM-Reference-Format}
\bibliography{main}

\appendix